\documentclass[aps,prx,reprint,groupedaddress]{revtex4-2}

\usepackage{graphicx}
\usepackage[dvipsnames]{xcolor}
\usepackage[colorlinks=true,linkcolor=Blue,citecolor=Blue,urlcolor=Blue,linktocpage=true]{hyperref}
\usepackage{amsmath,amssymb,amscd,amsfonts,bm, bbm}
\usepackage{verbatim}

\newcommand{\inpt}{\ensuremath{_{\rm in}}}
\newcommand{\out}{\ensuremath{_{\rm out}}}
\newcommand{\avg}[1]{\ensuremath{\langle #1\rangle}}
\newcommand{\diag}{\ensuremath{\rm diag}}

\renewcommand{\Re}{\ensuremath{{\rm Re}\,}}
\renewcommand{\Im}{\ensuremath{{\rm Im}\,}}

\begin{document}
	
	\title{Characterization of Linear Measurements in Cavity Optomechanics:\texorpdfstring{\\}{} Examples and Applications}
	
	\author{F. Bemani}
	\email{foroudbemani@gmail.com}
	\affiliation{Department of Optics, Palack\'{y} University, 17. listopadu 1192/12, 77900 Olomouc, Czechia}
	
	\author{O. \v{C}ernot\'{i}k}
	\email{ondrej.cernotik@upol.cz}
	\affiliation{Department of Optics, Palack\'{y} University, 17. listopadu 1192/12, 77900 Olomouc, Czechia}
	
	\author{R. Filip}
	\email{filip@optics.upol.cz}
	\affiliation{Department of Optics, Palack\'{y} University, 17. listopadu 1192/12, 77900 Olomouc, Czechia}
	
	\date{\today}
	
	\begin{abstract}
		Detailed understanding of physical measurements is essential for devising efficient metrological strategies and measurement-feedback schemes, as well as finding fundamental limitations on measurement sensitivity. In the quantum regime, measurements modify the state of the system of interest through measurement backaction as a direct consequence of the Heisenberg principle. In cavity optomechanics and electromechanics, a plethora of strategies exist for measuring the mechanical motion using electromagnetic fields, each leading to different competition between measurement imprecision and backaction. While this range of techniques allows broad applications of optomechanical and electromechanical devices, it makes direct comparison of different measurement methods difficult. We develop a formalism for quantifying the performance of optomechanical measurements using a few relevant figures of merit. Our approach is inspired by similar characterizations in quantum optics and quantifies the main properties of quantum measurements---the ability to distinguish different quantum states and preservation of signal in the presence of measurement noise. We demonstrate our concept on the most common optomechanical measurements---displacement detection, coherent quantum noise cancellation, and quantum nondemolition measurements---and perform detailed analysis of errors in optomechanical nondemolition measurements. This newly acquired knowledge allows us to propose a strategy for quantum nondemolition measurements in levitodynamics using coherent scattering. Our results complement existing knowledge of linear optomechanical interactions and open the way to new understanding of optomechanical measurements, thus allowing also novel applications of optomechanical devices in fundamental physics and quantum technologies.
	\end{abstract}
	
	\maketitle

	\section{Introduction}\label{sec:intro}
	
	Cavity optomechanics and electromechanics reached a remarkable level of controlling mechanical resonators using electromagnetic fields, all the way down to the quantum level~\cite{Aspelmeyer2014,Barzanjeh2022}. From pioneering experiments establishing the basic elements of optomechanical interactions~\cite{Fabre1994,Cohadon1999}, the field quickly matured to experiments demonstrating efficient quantum state preparation~\cite{Pirkkalainen2015,Wollman2015,Ockeloen-Korppi2018}, complex multimode behavior including topological effects~\cite{Lee2015,Xu2016,Nielsen2017}, advanced nonlinear dynamics~\cite{Sankey2010,Brawley2016,Navarro-Urrios2017}, and pivotal applications in quantum technologies~\cite{Barzanjeh2022,Midolo2018}. Hybridizing mechanical oscillators with other quantum systems further enhances their functionality, making mechanical devices an indispensable component of both fundamental quantum physics investigations and practical quantum technology applications~\cite{Chu2020,Clerk2020,Heinrich2021}.
	
	In most optomechanical and electromechanical platforms, the motion of the mechanical oscillator modifies the resonance frequency of an optical or microwave resonator, resulting in the radiation-pressure interaction (setting $\hbar = 1$)
	\begin{equation}
		H = \omega_c a^\dagger a + \omega_m b^\dagger b + g_0 a^\dagger a(b+b^\dagger),
	\end{equation}
	where $a$ and $b$ are the annihilation operators for the electromagnetic and mechanical mode, $\omega_c$ and $\omega_m$ their bare frequencies, and $g_0$ is the optomechanical coupling strength which depends on the particular implementation. Although the radiation-pressure interaction is always present, the single-photon optomechanical coupling $g_0$ is typically small due to the limited mechanical zero-point motion, resulting in small dispersive shift for quantum fluctuations. In addition, the large separation between electromagnetic and mechanical frequencies suppresses resonant energy exchange. The common solution is to drive the electromagnetic resonator with a classical pump which results in a strong quasiclassical amplitude $\alpha$ in the resonator. The interaction can then be linearized around this amplitude,
	\begin{equation}
		H = \Delta a^\dagger a + \omega_m b^\dagger b + g(a+a^\dagger)(b+b^\dagger).\label{Eq:Lin_H}
	\end{equation}
	Here, $\Delta = \omega_c - \omega_p$ is the detuning between the cavity resonance and the frequency of the external pump $\omega_p$, $g = g_0\alpha$, and we assumed $\alpha\in\mathbb{R}$ for simplicity.
	
	The detuning of the pump field from the cavity resonance is an important control parameter. Driving the system on the red mechanical sideband, $\Delta = \omega_c-\omega_p = \omega_m$, resonantly enhances the beam-splitter interaction $H_{\rm BS} = g(a^\dagger b+b^\dagger a)$ which can be used to cool the mechanical motion to its quantum ground state~\cite{Chan2011,Teufel2011,Peterson2016}, store a quantum state in a mechanical quantum memory~\cite{Palomaki2013,Reed2017}, or convert quantum signals between fields at different frequencies~\cite{Lecocq2016,Forsch2020,Mirhosseini2020,Brubaker2022,Delaney2022}. The efficiency of these tasks relies on the resolved-sideband regime where the cavity linewidth is much smaller than the mechanical frequency, $\kappa\ll\omega_m$, allowing the two-mode squeezing part of the interaction, $H_{\rm TMS} = g(ab+a^\dagger b^\dagger)$, to be neglected under the rotating wave approximation; however, a range of techniques exists that allow efficient beam-splitter coupling also for large sideband ratios $\kappa/\omega_m\gtrsim 1$~\cite{Elste2009,Jockel2015,Sawadsky2015,Clark2017,Rossi2017,Cernotik2019,Lau2020}. In contrast, driving the system on the blue sideband, $\Delta = \omega_c-\omega_p = -\omega_m$ enhances the two-mode squeezing interaction $H_{\rm TMS}$, leading to amplification~\cite{Ockeloen-Korppi2016a,Toth2017,Shen2018,MercierDeLepinay2019}, entanglement between the mechanical and cavity modes~\cite{Palomaki2013a,Riedinger2018}, and conditional generation of single phonons~\cite{Riedinger2016,Hong2017}.
	
	When the cavity is driven on resonance, $\Delta = \omega_c-\omega_p = 0$, the optical output field acquires a phase shift relative to the drive field that depends on the position of the mechanical oscillator, allowing its measurement in the unresolved sideband regime \cite{Regal2008,Anetsberger2009}. The accuracy of this displacement detection is limited by the standard quantum limit (SQL) which represents a compromise between the imprecision noise (which limits the signal-to-noise ratio of the detector) and backaction noise (which disturbs the mechanical state)~\cite{Regal2008,Anetsberger2009,Clark2016}. Various measurement strategies have been proposed and realized that limit or completely avoid the backaction noise, for example, measuring a single mechanical quadrature~\cite{Clerk2008,Hertzberg2009,Suh2014,Shomroni2019,Brunelli2019}, cancelling the effect of measurement backaction using interference~\cite{Tsang2010,Wimmer2014}, modifying the homodyne detection of the optical output~\cite{Buchmann2016,Kampel2017,Ockeloen-Korppi2018,Mason2019}, or engineering quantum mechanics free subspaces~\cite{Tsang2012,MercierDeLepinay2021}. The former two methods are especially powerful: Quantum nondemolition (QND) measurements, employing the rotating-frame Hamiltonian $H_{\rm QND} = g(a+a^\dagger)(b+b^\dagger)$ (which can be engineered by driving both mechanical sidebands with equal strengths~\cite{Clerk2008}), have been used for quantum tomography of nonclassical mechanical states~\cite{Lecocq2015,Lei2016,Ockeloen-Korppi2016} and are a powerful tool also in quantum optics~\cite{Levenson1986,LaPorta1989} and atomic physics~\cite{Kuzmich2000,Appel2009,Sewell2013}. Coherent quantum noise cancellation (CQNC) employs an auxiliary oscillator with negative effective mass which ensures that the effects of measurement backaction on the mechanical oscillator and the negative-mass oscillator cancel out, allowing not only measurements free of backaction~\cite{Moller2017} but also more efficient generation of entanglement~\cite{Huang2018,Thomas2021}.  In addition to continuous measurement schemes, stroboscopic quantum optomechanical driving can be used to achieve ground-state cooling, squeezing, and backaction-evading measurements \cite{Brunelli2020}.
	
	While the beam-splitter and two-mode squeezing interactions are well understood, detailed characterization and comparison of different measurement strategies is still missing. Given the importance of efficient measurements for reconstructing quantum states and processes and verifying their properties~\cite{Genoni2015,Gut2020}, estimating external forces and fields in quantum metrology~\cite{Schreppler2014,Kampel2017,Mason2019}, and implementing measurement-based feedback control~\cite{Wilson2015,Schafermeier2016,Rossi2018}, detailed understanding of available strategies is crucial for selecting the optimal approach for any given task. In addition, such knowledge would complement existing theory for beam-splitter and two-mode squeezing interactions and complete our understanding of the linear regime of cavity optomechanics and electromechanics.
	
	Here, we apply characterization techniques from quantum optics~\cite{Roch1992,Grangier1998} on optomechanical  measurements and evaluate their performance by their ability to generate squeezed mechanical states and by the efficiency of transferring the mechanical signal into the optical and mechanical outputs. These quantifiers were originally developed for analysing QND measurements in quantum optics where they are crucial to reduce backaction noise of linear photonic elements in measurements~\cite{Bruckmeier1997} or quantum operations~\cite{Bowen2003,Hetet2008}. We extend their use beyond photonic devices to cavity optomechanical and electromechanical systems and beyond analysis of QND measurements. In particular, we show that, unsurprisingly, conventional displacement detection is a fully classical measurement with optimal performance attained at the SQL; however, detecting the signal away from mechanical resonance allows suppression of measurement backaction, which leads to increased transfer of the signal to the mechanical output. Moreover, we demonstrate that CQNC is, remarkably, also fully classical when measuring on mechanical resonance but it can reach the QND regime with off-resonant detection.
	
	For optomechanical single-quadrature measurements, we find a simple and experimentally accessible condition on optomechanical cooperativity to reach the QND regime in the ideal case. This condition depends only on the initial mechanical variance and---for electromechanical systems affected by thermal noise in the microwave input---temperature of the electromagnetic bath. Afterwards, we analyze the role of various imperfections on the quality of the QND readout. We consider not only counter-rotating terms in the optomechanical interaction but also general linear mechanical dynamics and free cavity oscillations (the intrinsic oscillation of the cavity field in the form $\nu a^\dagger a$ with frequency $\nu$). These additional contributions to the total Hamiltonian can stem from a nonzero detuning of the optical driving tones from the mechanical sidebands (giving rise to free oscillations of the mechanical or optical mode) or nonlinearities of the mechanical dynamics (resulting in mechanical squeezing). We show that it is possible to use one to compensate the other (e.g., intentionally introduce detuning to account for intrinsic mechanical squeezing) and obtain perfect QND readout even in the presence of these systematic imperfections.
	
	Finally, we use this newly acquired knowledge to propose and analyze a strategy for QND measurement of the motion of a levitated nanoparticle using coherent scattering~\cite{Gonzalez_Ballestero2019}. In this system, modulation of a tweezer amplitude at the mechanical frequency allows strong QND-like coupling of a single mechanical quadrature to an optical cavity mode. However, coherent scattering of photons between the two mechanical sidebands (in which phonons are created or annihilated in pairs) leads to additional mechanical squeezing~\cite{Cernotik2020} which destroys the QND nature of the readout. Suitable detuning of the modulation frequency from the exact mechanical resonance can then be used to obtain true QND measurement of the nanoparticle motion, opening new avenues to tests of fundamental physics~\cite{Moore2021} and quantum technology applications in levitodynamics~\cite{Gonzalez-Ballestero2021}.
	Backaction-evading measurements together with feedback can generate unconditional mechanical squeezing in optomechanical systems, offering an example where optimizing measurements through systematic characterization leads to tangible quantum control advances \cite{DiGiovanni2021}.

	\section{Measurement characterization}\label{sec:TVdiagram}

	\subsection{System dynamics and QND criteria}\label{ssec:model}
	
\begin{figure}
		\centering
		\includegraphics[width=\linewidth]{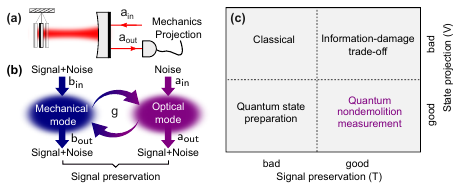}
		\caption{\label{fig:schematic} Characterization of linear optomechanical measurements. (a) Schematic depiction of an optomechanical setup with optical input $a\inpt$ and output $a\out$. The output is measured using homodyne detection which allows us to infer the mechanical state. (b) The cavity and mechanical resonator interact at a rate $g$ (enhanced by driving the cavity) which transfers the input signal into both optical and mechanical output. Each output contains also noise contributions from the cavity and mechanical inputs; in addition, each mode might have its own free dynamics which mix its quadrature operators. Comparing the strength of the signal to the noise enables us to quantify the preservation of the signal by the measurement. (c) Natural bounds on the quality of state inference $V$ (mechanical variance below the shot noise level) and signal transfer $T$ (larger than unity) allows us to define four possible measurement regimes.}
	\end{figure}

	The main principles of our approach to characterizing optomechanical measurements are illustrated in Fig.~\ref{fig:schematic}. Cavity optomechanical interaction is used to imprint information about the motion of a mechanical resonator to an optical field [Fig.~\ref{fig:schematic}(a)]. The cavity output is then measured in a homodyne measurement which projects the mechanical resonator to a certain state; ideally, this measurement ought to prepare a squeezed mechanical state. Formally, the measurement tries to estimate an unknown signal in the presence of mechanical (thermal) and optical (measurement) noise; see Fig.~\ref{fig:schematic}(b). The key features of the measurement are then the information about the mechanical displacement and the preservation of the signal for repeated measurements, which can be quantified using the conditional mechanical variance and signal transmission coefficients and plotted jointly in a so-called $TV$ diagram [Fig.~\ref{fig:schematic}(c)]. Comparison with natural bounds separates this phase diagram into four regions corresponding to different measurement regimes which we will describe in detail below.
	
	To model and characterize optomechanical measurements, we start from the Hamiltonian
	\begin{equation}\label{eq:DispDet}
		H = \omega_m b^\dagger b + g(a+a^\dagger)(b+b^\dagger),
	\end{equation}
	which describes usual displacement detection using a resonant optical pump~\cite{Wilson2015,Schafermeier2016}. In the following, we describe the modes in terms of their dimensionless quadrature operators, $X = (a+a^\dag)/\sqrt{2}$, $Y = -i(a-a^\dag)/\sqrt{2}$ for the cavity field and $x = (b+b^\dag)/\sqrt{2}$, $p=-i(b-b^\dag)/\sqrt{2}$ for the mechanical modes, satisfying $[X,Y]=[x,p]=i$. The equations of motion with dissipation and noise included then read~\cite{Aspelmeyer2014}
	\begin{subequations}\label{eq:EOM}
		\begin{align}
			\dot X &= -\frac{\kappa}{2}X + \sqrt{\kappa} X\inpt,\\
			\dot Y &= -2gx -\frac{\kappa}{2}Y + \sqrt{\kappa} Y\inpt,\\
			\dot x &= \omega_m p -\frac{\gamma}{2}x + \sqrt{\gamma} x\inpt,\\
			\dot p &= -\omega_m x - 2gX -\frac{\gamma}{2}p + \sqrt{\gamma} p\inpt,
		\end{align}
	\end{subequations}
	where $\kappa,\gamma$ are the optical and mechanical decay rates.
	The operators  $X\inpt,Y\inpt,x\inpt,p\inpt$ describe the noise entering the optical and mechanical modes and their correlations satisfy \cite{Wiseman2010}
	\begin{subequations}\label{eq:cav_in}
		\begin{align}
			\langle X_{\rm in}(t)X_{\rm in}(t')\rangle &= \langle Y_{\rm in}(t)Y_{\rm in}(t')\rangle  = \frac{1}{2}\delta (t - t'),\\
			\langle X_{\rm in}(t){Y_{\rm in}}(t')\rangle &=  - \langle Y_{\rm in}(t)X_{\rm in}(t')\rangle  =   \frac{i}{2}\delta (t - t'),
		\end{align}
	\end{subequations}
	for the optical inputs and
	\begin{subequations}\label{eq:minput}
		\begin{align}
			\left\langle x_{\rm in } (t)x_{\rm in } (t') \right\rangle  &= \left(n_m + \Re m_{\rm sq} + \frac{1}{2} \right)\delta (t - t'),\\
			\left\langle p_{\rm in } (t)p_{\rm in } (t') \right\rangle  &= \left(n_m - \Re m_{\rm sq} + \frac{1}{2} \right)\delta (t - t'),\\
			\langle x_{\rm in } (t)p_{\rm in } (t')\rangle  &= \left(\frac{i}{2} + \Im m_{\rm sq} \right)\delta (t - t'),\\
			\langle p_{\rm in } (t)x_{\rm in } (t')\rangle  &= \left(-\frac{i}{2} + \Im m_{\rm sq} \right)\delta (t - t'),
		\end{align}
	\end{subequations}
	for the mechanical noise operators; all other correlations are zero. Here $n_m$ is the mean occupation of the mechanical bath and $m_{\rm sq}$ parametrizes its squeezing; the two parameters satisfy $|m_{\rm sq}|^2\leq n_m(n_m+1)$ with equality attained for a bath in a pure squeezed state~\cite{Wiseman2010}. The correlation functions~\eqref{eq:minput} describe a general thermal squeezed bath for the mechanical resonator. Measurement according to Eq.~\eqref{eq:DispDet} can thus be used to characterize such a bath or perform tomography of a dissipatively prepared squeezed mechanical state~\cite{Wollman2015,Pirkkalainen2015}. The measurement can easily be extended to estimation of external forces or fields by appropriately modifying the correlation functions.
	
	The equations of motion~\eqref{eq:EOM} are complemented with the usual input--output relations for the optical and mechanical quadratures, $X_{\rm out} = \sqrt{\kappa}X - X_{\rm in}$, $x_{\rm out} = \sqrt{\gamma}x-x_{\rm in}$ (with analogous expressions for $Y_{\rm out}$ and $p_{\rm out}$). Note that the mechanical output is not accessible in an experiment as it describes the mechanical reservoir. However, including it in the mathematical description allows us to treat the optical and mechanical mode on an equal footing and directly apply the techniques used in quantum optics where the output of the measured mode is available in general. Since the mechanical output is directly proportional to the mechanical mode quadratures $x,p$, it can be used to fully characterize the state of the mechanical mode.
	
	To solve the system dynamics and characterize the measurement, we define the vector of quadrature operators ${\bf u} = (X,Y,x,p)^T$ (with vectors of input and output noise operators ${\bf u}_{\rm in,out}$ defined similarly). We then solve Eqs.~\eqref{eq:EOM} in the frequency space defined via the Fourier transform $o(\omega)=\int o(t) e^{i \omega t}dt$. The output quadratures are then related to the input via the scattering matrix ${\bf S}(\omega)$ (see Appendix~\ref{app:efficiency}), 
	\begin{equation} \label{eq:SM}
		{\bf u}_{\rm out}(\omega) = {\bf S}(\omega){\bf u}_{\rm in}(\omega).
	\end{equation}
	For the output quadratures, we can define the covariance matrix with elements
	\begin{align} 
		\begin{split}
			V_{{\rm{out}}}^{ij}(\omega)\delta(\omega+\omega') &= \frac{1}{2}\avg{u_{\rm out}^i(\omega)u_{\rm out}^j(\omega') + u_{\rm out}^j(\omega)u_{\rm out}^i(\omega')} \\
			&\quad- \avg{u_{\rm out}^i(\omega)}\avg{u_{\rm out}^j(\omega')}
		\end{split}
	\end{align}
	which contains information about all correlations between the mechanical and optical modes and can be obtained as 
	\begin{equation}
		{\bf V}_{\rm out}(\omega) = \frac{1}{2}[{\bf S}(\omega){\bf V}_{\rm in}{\bf S}^T(-\omega) + {\bf S}(-\omega){\bf V}_{\rm in}{\bf S}^T(\omega)],
	\end{equation}
	where the covariance matrix of the input operators is defined analogically and can be obtained from the correlation functions~\eqref{eq:cav_in}, \eqref{eq:minput},
	\begin{equation}\label{eq:Vin}
		{\bf V}_{\rm in} = \begin{pmatrix}
			\frac{1}{2} & 0 & 0 & 0 \\
			0 & \frac{1}{2} & 0 & 0 \\
			0 & 0 & n_m + \Re m_{\rm sq} + \frac{1}{2} & \Im m_{\rm sq} \\
			0 & 0 & \Im m_{\rm sq} & n_m - \Re m_{\rm sq} + \frac{1}{2} \\
		\end{pmatrix}.
	\end{equation}
	
	We can now characterize the quality of the measurement borrowing techniques for evaluating QND measurements in quantum optics~\cite{Roch1992,Grangier1998}. These evaluations are based on three quantities:
	First, the conditional mechanical variance $V_c$ quantifies the ability of the measurement to generate mechanical squeezing and thus specifies the measurement resolution. Mathematically, it is given by
	\begin{equation}\label{eq:Vc}
		V_c = V_{\rm out}^{33} - \frac{|V_{\rm out }^{32}|^2}{V_{\rm out }^{22}},
	\end{equation}
	where the output covariance matrix is evaluated at the frequency of the readout. Here, $V_{33}$ and $V_{22}$ represent the variances of the mechanical position and optical phase quadratures, respectively, while $V_{32}$ denotes their covariance. The conditional variance $V_c$ represents the uncertainty of the mechanical position quadrature $x$ after optimal linear inference based on the available measurement outcomes, i.e., the optical output quadrature $Y^{\rm out}$. A genuine quantum measurement has to be able to reach resolution below the vacuum variance of $\frac{1}{2}$ so we require $V_c < \frac{1}{2}$ for a QND measurement or any other quantum-limited measurement.
	
	Second, the signal transfer coefficient $T_s$ characterizes transfer of the initial mechanical signal from the mechanical input to the mechanical output. It can be calculated as
	\begin{equation}\label{eq:Ts}
		T_s = \frac{{V_{\rm in}^{33}}}{{V_{\rm in}^{33} + n_s^{\rm eq}}}, \qquad n_s^{\rm eq} = \frac{V_{\rm out}^{33}}{|S_{33}|^2} - V_{\rm in}^{33},
	\end{equation}
	where we introduced the measurement-equivalent input noise $n_s^{\rm eq}$ in terms of the variances $V_{\rm in,out}^{33}$ and the scattering matrix element $S_{33}$. Third, the meter transfer coefficient $T_m$ represents the amount of information about the signal transferred to the cavity output and can be obtained in the same way as the signal transfer coefficient,
	\begin{equation}\label{eq:Tm}
		T_m = \frac{{V_{\rm in}^{33}}}{{V_{\rm in}^{33} + n_m^{\rm eq}}}, \qquad n_m^{\rm eq} = \frac{V_{\rm out}^{22}}{| S_{23}|^{2}} - V_{\rm in}^{33}.
	\end{equation}
	Nondemolition measurements must have both transfer coefficients nonzero. Without the signal being preserved in the system (quantified here by its propagation to system output to keep symmetry with the meter), the measurement cannot be repeated; without signal transfer to the meter, there is no measurement. We therefore require $T_s+T_m > 1$ for QND measurements. On the other hand, classical measurements satisfy the bound $T_s + T_m \le 1$ in agreement with the standard quantum limit. A strong classical measurement extracts a lot of information (high $T_m$) at the expense of introducing measurement backaction (low $T_s$), whereas preserving the signal in the system of interest (high $T_s$) leads to weak information extraction by the meter (low $T_m$)~\cite{Roch1992,Grangier1998}.

	\subsection{Operational meaning of measurement regimes}\label{ssec:meaning}
	The two thresholds, $V_c = \frac{1}{2}$ and $T_s + T_m = 1$, define four distinct measurement regimes, following the standard classification of quantum nondemolition (QND) measurements developed in quantum optics~\cite{Grangier1998,Sewell2013}, see Fig.~\ref{fig:schematic}(c). The boundary $V_c = \frac{1}{2}$ corresponds to the vacuum (standard quantum limit) variance of a single mechanical quadrature and therefore separates measurements capable of conditional squeezing from those limited to classical resolution. The bound $T_s + T_m = 1$ represents the classical information--disturbance limit: measurements satisfying $T_s + T_m \le 1$ necessarily trade signal preservation for information gain, whereas violations of this bound are a defining signature of QND measurements enabled by the commutativity of the measured observable at different times.
	
	When both conditions are satisfied simultaneously, $V_c < \frac{1}{2}$ and $T_s + T_m > 1$, the measurement operates in the genuine QND regime. In this case, information about the mechanical quadrature is extracted with sub-vacuum resolution while the quadrature itself is preserved, enabling simultaneous quantum state preparation and repeated or continuous monitoring. The ideal QND limit, $V_c \to 0$ and $T_s + T_m \to 2$, corresponds to projection onto mechanical quadrature eigenstates with perfect signal preservation.
	
	In the opposite regime, $V_c \ge \frac{1}{2}$ and $T_s + T_m \le 1$, the measurement is fully classical. The first condition specifies that the measurement resolution is insufficient to reduce mechanical variance below the vacuum level; the second condition signifies that we cannot obtain meaningful information about the mechanical quadrature and preserve its state for future measurements. Such measurements are fundamentally limited by the standard quantum limit.
	
	The remaining two regimes correspond to intermediate quantum measurement scenarios. For $V_c < \frac{1}{2}$ but $T_s + T_m \le 1$, the measurement can conditionally generate mechanical squeezing, but does so at the expense of signal demolition; this regime is therefore useful for single-shot quantum state preparation but is not practical for repeated measurements. Conversely, when $V_c \ge \frac{1}{2}$ and $T_s + T_m > 1$, the measurement preserves the mechanical observable and establishes strong correlations between the mechanical resonator and the optical meter, but without sufficient resolution to surpass the vacuum noise level. This regime highlights the information--disturbance tradeoff intrinsic to weak but nondestructive quantum measurements.

	\subsection{Characterizing measurements based on experimental data}
	
	The primary experimental observable through which all calibration and experimental data is collected is the homodyne photocurrent proportional to the optical phase quadrature $Y_{\rm out}$. From this signal, the symmetrized power spectral density $S_{YY}(\omega)$ is obtained. After correcting for the independently calibrated detection efficiency and electronic noise, this spectrum directly yields the optical output variance $V_{\rm out}^{22}(\omega)$ within our normalization. For the \emph{TV} diagram formalism to be practical, the conditional mechanical variance $V_c$ and signal transfer coefficients $T_{s,m}$ must be attainable from this data as well.
	
	The system parameters---optical and mechanical decay rates $\kappa$, $\gamma$, mechanical resonance frequency $\omega_m$, linearized optomechanical coupling strength $g$, and mechanical bath occupation $n_m$---are extracted by fitting the measured optical noise spectrum to the theoretical response derived from Eqs.~\eqref{eq:DispDet}--\eqref{eq:SM}. The thermal occupation $n_m$ provides absolute calibration obtained by analyzing the mechanical sideband area in the weak-coupling regime followed by standard optomechanical calibration procedures. The full measurement characterization is then achieved by combining calibrated optical noise spectra, independently determined system parameters, and linear-response theory to reconstruct the relevant output covariance matrix elements.
	
	The conditional mechanical variance $V_c$ is not a directly measurable observable but is inferred from the experimentally determined output variances and correlations. Specifically, $V_c$ is computed from the optical phase-quadrature variance $V_{\rm out}^{22}$, mechanical position-quadrature variance $V_{\rm out}^{33}$, and their inferred covariance $V_{\rm out}^{32}$ using Eq.~\eqref{eq:Vc}. This procedure relies on the assumption of linear dynamics and Gaussian noise and quantifies the mechanical state prepared by conditioning on the homodyne measurement record. As the mechanical output channel is not directly accessible, the signal transfer coefficient $T_s$ is inferred within the linear Gaussian model using the calibrated mechanical susceptibility and the optically induced backaction noise, which is determined by the known coupling strength $g$ and cavity decay rate $\kappa$. The equivalent input noise $n_s^{\rm eq}$ is then obtained from the inferred mechanical displacement-quadrature output variance $V_{\rm out}^{33}(\omega)$, yielding $T_s$ through Eq.~\eqref{eq:Ts}. Finally, the meter transfer coefficient $T_m$ is extracted from the calibrated optomechanical response function $S_{23}(\omega)$ and the measured optical noise spectrum $V_{\rm out}^{22}(\omega)$. From these quantities, the measurement-equivalent input noise $n_m^{\rm eq}$ is determined, allowing direct evaluation of $T_m$ via Eq.~\eqref{eq:Tm}.

	\section{Analysis of optomechanical measurements}

	\subsection{Displacement detection}\label{ssec:displacement}
	
\begin{figure}
		\centering
		\includegraphics[width=\linewidth]{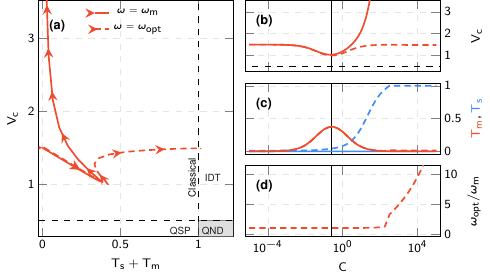}
		\caption{\label{fig:displ}
			$TV$ diagram for displacement detection. (a) Conditional variance $V_c$ plotted against the sum of the transfer coefficients $T_s+T_m$ as the optomechanical cooperativity is increased from $C= 0$ to $C\to\infty$ (shown by the arrows). Solid line corresponds to detection on mechanical resonance, $\omega = \omega_m$, while the dashed line is for a numerically optimized detection frequency $\omega = \omega_{\rm opt}$ which gives the minimum conditional variance for a given cooperativity. The vertical dashed line shows the bound $T_s+T_m = 1$; the horizontal dashed line gives the vacuum variance of $\frac{1}{2}$. Together, they separate the parameter space into four quadrants: classical regime (upper left), information--damage tradeoff (IDT, upper right), quantum state preparation (QSP, lower left), and QND readout (lower right). (b) Conditional variance $V_c$, (c) transfer coefficients $T_s$ (blue), $T_m$ (red), and (d) optimal detection frequency [corresponding to the dashed lines in panels (a)--(c)] as functions of the optomechanical cooperativity. The solid vertical lines show the cooperativity corresponding to the SQL. The system parameters are $\kappa/\omega_m = 10$, $\gamma/\omega_m = 0.01$, $n_m=1$, and $m_{\rm sq}=0$.}
	\end{figure}

	We illustrate this behavior for displacement detection in Fig.~\ref{fig:displ}. Panel (a) shows the $TV$ diagram of this measurement, in which the conditional variance $V_c$ is plotted against the sum of the transfer coefficients $T_s+T_m$ (all evaluated on mechanical resonance, $\omega = \omega_m$) as the optomechanical cooperativity $C = 4 g^2/\kappa\gamma$ is increased from $C=0$ to $C\to\infty$. The two conditions on QND measurements split the diagram into four quadrants corresponding to the four possibilities discussed above. In the limit $C\to 0$, no measurement takes place and we have no signal transfer to the cavity output, $T_m = 0$, and conditional variance equal to the initial variance, $V_c = V_x = n_m+\Re m_{\rm sq}+\frac{1}{2}$; in addition, $T_s\approx 0$ as well, since $n_s^{\rm eq}\simeq V_x/|S_{33}|^2\simeq 16\omega_m^2V_x/\gamma^2\ggg 1$ on mechanical resonance due to mixing of both mechanical quadratures. s the cooperativity is increased, the conditional variance slowly decreases and the transfer coefficient $T_m$ increases up to the SQL. Afterwards, the measurement is dominated by backaction noise, causing the conditional variance to diverge, and the transfer coefficients decrease back to zero for $C\to\infty$. Throughout the trajectory in the $TV$ diagram, the displacement detection remains in the classical regime.
	
	In Fig.~\ref{fig:displ} (b) and (c), we plot the conditional variance $V_c$ and transfer coefficients $T_{s,m}$ versus the optomechanical cooperativity $C$ to analyze the measurement in more detail. The conditional variance [Fig. \ref{fig:displ}(b)] gradually decreases at first until it reaches the cooperativity corresponding to the SQL, $C_{\rm SQL}\approx \frac{1}{4} + \omega_m^2/\kappa^2$ (denoted by the vertical line; see Appendix~\ref{app:efficiency}). Beyond this cooperativity, the conditional variance starts diverging due to the measurement backaction. Moreover, while the transfer coefficient $T_s$ remains close to zero for any cooperativity (owing to the strong backaction affecting the measured position quadrature), the coefficient $T_m$ reaches its maximum at $C_{\rm SQL}$ [Fig. \ref{fig:displ}(c)]. In the limits $C\to 0$ and $C\to\infty$, the measurement sensitivity is limited by, respectively, the imprecision and backaction noise, which reduce the transfer coefficients; only when both are minimized simultaneously (close to the SQL), non-negligible transfer of signal to the output (as quantified by $T_m$) is possible.
	
	Measuring exactly on mechanical resonance is, however, not always optimal. This is shown in Fig.~\ref{fig:displ} by the dashed lines which correspond to minimizing the conditional variance for given value of cooperativity over the detection frequency $\omega$. When the measurement is limited by imprecision noise ($C<C_{\rm SQL}$), the optimal measurement is on mechanical resonance, $\omega_{\rm opt}\approx\omega_m$. On the other hand, off-resonant detection (with $\omega_{\rm opt} > \omega_m$) is ideal when the measurement is limited by backaction noise ($C > C_{\rm SQL}$). The conditional variance then saturates at the initial mechanical variance $V_x$ in the limit $C\to\infty$ instead of diverging; for large cooperativity, $C\gg C_{\rm SQL}$, the transfer coefficient $T_s$ approaches unity and the coefficient $T_m$ keeps a small nonzero value, shifting the measurement to the IDT regime. This behavior remains qualitatively the same for any initial mechanical variance.
	
	With limited detection efficiency $\eta\in(0,1)$ caused by optical losses in the measurement chain, the transfer coefficients and conditional variance are affected with some terms (though not all) acquiring renormalization of the optomechanical cooperativity, $C\to C\eta$, as we derive in Appendix~\ref{app:efficiency}. While these effects are nontrivial, they are easy to understand: The signal transfer coefficient $T_s$ depends on the optomechanical cooperativity $C$ but not on the detection efficiency $\eta$ as the total measurement backaction depends only on the total number of photons interacting with the mechanical resonator and not on whether we detect them after the interaction or not; on the other hand, the meter transfer coefficient $T_m$ sees the renormalization $C\to C\eta$ as detection losses affect how much we learn about the mechanical state. Similarly, in the conditional variance $V_c = V_{\rm out}^{33} - {|V_{\rm out }^{32}|^2}/{V_{\rm out }^{22}}$, the first term is unaffected by detection efficiency since it represents the unconditional variance of the mechanical mode, while the second term quantifies the acquired knowledge about the mechanical state and the cooperativity is here again renormalized by the detection efficiency.

	\subsection{Coherent quantum noise cancellation}\label{ssec:CQNC}
	
	The effect of backaction noise can be reduced by introducing an auxiliary system that has equal but opposite response to the input optical quadrature $X\inpt$ as the mechanical resonator. In the output phase quadrature $Y\out$, the contributions of the measurement backaction noise $X\inpt$ from the mechanical mode $b$ and this additional mode $c$ interfere destructively, leading to displacement measurement unaffected by measurement backaction. It can be shown that such response can be achieved with a mode that has the same properties as the mechanical mode (same coupling strength to the optical cavity, frequency, and linewidth) but effective negative mass \cite{Tsang2010}. Such physics can be achieved in an atomic ensemble~\cite{Moller2017} or using an additional optical cavity~\cite{Wimmer2014}. 
	
	We now rewrite the linearized optomechanical Hamiltonian in terms of
	dimensionless quadrature operators as
	\begin{equation}
		H_0 = \frac{\omega_m}{2}\left(x^2+p^2\right) + 2gXx .
	\end{equation}
	To implement coherent quantum noise cancellation, we introduce an auxiliary mode $c$ with effective negative mass and identical coupling strength.
	The total Hamiltonian is then extended to
	\begin{equation}\label{eq:H_CQNC}
		H = \frac{\omega_m}{2}\left(x^2+p^2-X_c^2-Y_c^2\right)
		+ 2gXx + 2gXX_c ,
	\end{equation}
	where $X_c,Y_c$ are the quadratures of the auxiliary mode $c$, defined in full analogy to $X,Y$.
	We can derive the scattering matrix in the same way as for the displacement detection (see Appendix~\ref{app:cqnc}) and evaluate the conditional variance and transfer coefficients.
	
	\begin{figure}
		\centering
		\includegraphics[width=\linewidth]{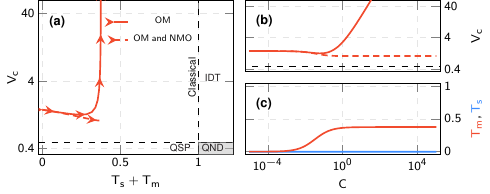}
		\caption{\label{fig:CQNC}
			Characterization of coherent quantum noise cancellation with detection on mechanical resonance, $\omega = \omega_m$. (a) $TV$ diagram showing the conditional variance $V_c$ versus the sum of the transfer coefficients $T_s+T_m$. The solid line shows the variance conditioned on the optical output only as described by Eq.~\eqref{eq:Vc}; the dashed line is for conditioning on both the optical and negative-mass mode, Eq.~\eqref{eq:Vc_CQNC}. (b) Conditional variance $V_c$ and (c) transfer coefficients $T_{s,m}$ against cooperativity. Parameters are the same as in Fig.~\ref{fig:displ} and we assume that the negative-mass oscillator has the same initial variance as the mechanical resonator.}
	\end{figure}
	
	The $TV$ diagram for CQNC with detection on mechanical resonance, $\omega = \omega_m$, is shown in Fig.~\ref{fig:CQNC}. The main difference from displacement detection is that the meter transfer coefficient $T_m$ retains a moderately large, nonzero value for cooperativities beyond the SQL (cf. Fig.~\ref{fig:displ}). The conditional variance still diverges for strong optomechanical cooperativity, but this is caused by strong quantum correlations between the mechanical resonator and the negative-mass oscillator and not by measurement backaction. Since these correlations indicate that some of the information about the mechanical displacement quadrature $x$ is imprinted onto the negative-mass oscillator quadrature $X_c$, this divergence can be removed assuming a (hypothetical) measurement of $X_c$. Such a measurement would allow us to condition the mechanical variance on both the optical output and the negative-mass oscillator (dashed line), obtained by extending the definition of conditional variance, Eq.~\eqref{eq:Vc}, to
	\begin{equation}\label{eq:Vc_CQNC}
		V_c=V_{33}^{\rm out}-\frac{|V_{32}^{\rm out}|^2}{V_{22}} -\frac{|V_{35}^{\rm out}|^2}{V_{55}},
	\end{equation}
	where the quadrature vector has been extended to ${\bf u} = (X,Y,x,p,X_c,Y_c)^T$. Eq.~\eqref{eq:Vc_CQNC} assumes that the interaction of the mechanical resonator and negative-mass oscillator with the common cavity field as described by the Hamiltonian~\eqref{eq:H_CQNC} entangles the mechanical position quadrature $x$ predominantly with the negative-mass amplitude quadrature $X_c$, which we justify in Appendix~\ref{app:cqnc}.
	
	Crucially, CQNC remains in the classical regime for both cases of conditioning despite fully suppressing backaction noise in measurement. This is because, as Fig.~\ref{fig:CQNC} shows, only the meter transfer coefficient remains nonzero (providing sufficient signal-to-noise ratio at the detector) while the signal transfer coefficient $T_s$ remains close to zero. This is a natural consequence of the interference between backaction noise contributions after interactions with the mechanical resonator and negative-mass oscillator---perfect destructive interference of backaction noise is possible only at the optical output, resulting in efficient signal transfer in the strong cooperativity regime, $C>C_{\rm SQL}$ (below the SQL, the measurement is limited by imprecision noise, resembling standard displacement detection). On the other hand, the mechanical resonator (and particularly its position quadrature $x$) still experiences measurement backaction, which suppresses the signal transfer coefficient $T_s$. In addition, the conditional variance cannot drop below the shot-noise level due to the excess noise provided by the negative-mass oscillator.
	
	\begin{figure}
		\centering
		\includegraphics[width=\linewidth]{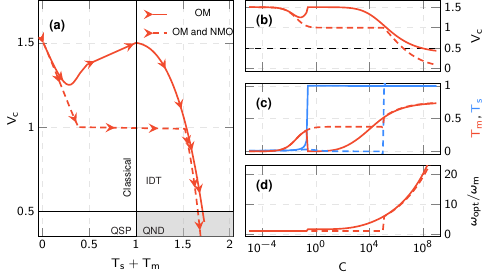}
		\caption{\label{fig:CQNC_opt}
			$TV$ diagram for optimal CQNC measurement. (a) Conditional variance $V_c$ optimized over measurement frequency plotted against the sum of the transfer coefficients $T_s+T_m$. (b) Conditional variance $V_c$, (c) transfer coefficients $T_{s,m}$, and (d) optimal detection frequency $\omega_{\rm opt}$ as functions of optomechanical cooperativity $C$. The meaning behind the different curves and system parameters are the same as in Fig.~\ref{fig:CQNC}.}
	\end{figure}
	
	Similar to displacement detection, measurement away from the mechanical resonance can lead to smaller conditional variance and larger transfer coefficients. We analyze this optimum numerically in Fig.~\ref{fig:CQNC_opt} which shows that off-resonant detection with sufficiently large optomechanical cooperativity is capable of reaching the QND regime, although not the ideal QND measurement with $V_c\to 0$, $T_s+T_m\to 2$. As before, we numerically optimize the conditional variance over detection frequency for given optomechanical cooperativity, for conditioning on the optical output only and for conditioning also on the negative-mass oscillator. Two surprising features can be seen in the $TV$ diagram in Fig.~\ref{fig:CQNC_opt}(a): First, conditioning on the optical output alone is sufficient to bring the mechanical variance below the vacuum level, albeit with prohibitively large optomechanical cooperativity $C\sim 10^8$. Second, there is a discontinuity in the transfer coefficients around $C\sim 10^6$ when conditioning on both optical output and negative-mass oscillator caused by a jump between two solution branches. For small to medium cooperativity, the measurement eventually saturates with $V_c\simeq 1$, $T_s = 0$, $T_m \simeq 0.4$ (remaining in the classical regime); for large cooperativity, the optimal measurement is close to conditioning on the optical measurement only, moving from IDT to QND regime. However, the relatively small optomechanical cooperativities in recent CQNC experiments, $C\lesssim 100$~\cite{Moller2017,Thomas2021}, is insufficient to reach the QND regime.
	
	Finally, we note that detection losses have the same effects as for displacement detection since the optomechanical interaction and subsequent measurement have the same form. The meter transfer coefficient and term in the conditional variance quantifying the gain in our knowledge of the mechanical state see renormalization of the cooperativity by the detection efficiency $C\to C\eta$ while the remaining terms remain unaffected (even though they depend on cooperativity). Similar features could also be expected for the (hypothetical) measurement of the negative-mass oscillator if it had its own detection inefficiency.

	\subsection{Quantum nondemolition measurement}\label{ssec:QND}
	
	In the following, we first evaluate the QND criteria assuming unit detection efficiency, $\eta=1$, in order to isolate the fundamental limits set by the interaction dynamics; the impact of finite detection efficiency, which is an important practical limitation, is discussed explicitly below. The QND criteria can be easily calculated for the ideal optomechanical QND readout described by the Hamiltonian~\cite{Clerk2008}
	\begin{equation}
		H_{\rm QND} = 2gXx.
	\end{equation}
	This form of interaction can be obtained with resonant driving ($\Delta = 0$) modulated at the mechanical frequency, resulting in $g(t) = g\cos(\omega_m t)$. In the interaction picture with respect to the free mechanical oscillations $\omega_m(x^2+p^2)$ and resolved sideband regime ($\kappa < \omega_m$), any fast-oscillating terms can be neglected under the rotating wave approxmation, the interaction reduces to an effective coupling between the optical amplitude quadrature $X$ and a single mechanical quadrature $x$, yielding the Hamiltonian $H_{\rm QND}$.
	
	This Hamiltonian and the corresponding equations of motion can be obtained from the displacement detection, Eq.~\eqref{eq:DispDet}, by taking the limit $\omega_m\to 0$ which allows us to directly evaluate the conditional variance and transfer coefficients,
	\begin{subequations}\label{eq:QNDIdeal}
		\begin{align}
			V_c &= \frac{1}{V_x^{-1} +32 C}, \label{Delta_x} \\
			T_s &= 1, \\
			T_m &= \frac{32C}{V_x^{-1} +32 C}. \label{Tm}
		\end{align}	
	\end{subequations}
	The conditional variance $V_c$ reduces below the vacuum level for cooperativity satisfying
	\begin{equation}\label{eq:C_bound}
		C > \frac{2V_x-1}{32V_x}
	\end{equation}
	which ranges from $C> 0$ for $V_x\leq\frac{1}{2}$ to $C > \frac{1}{16}$ for very noisy mechanical bath ($V_x\to\infty$). [The right-hand side of Eq.~\eqref{eq:C_bound} becomes negative for $V_x<\frac{1}{2}$ but the cooperativity is nonnegative by definition.] The sum of the transfer coefficients is greater than unity for any nonzero cooperativity, signifying efficient transfer of the signal to the meter and its preservation during the interaction; notably, we always have $T_s=1$ since the mechanical quadratures are not mixed by the QND dynamics. Moreover, by eliminating the cooperativity from Eqs.~\eqref{eq:QNDIdeal}, we obtain
	\begin{equation}\label{eq:QND_Vc}
		V_c = -(T_s+T_m-2)V_x
	\end{equation}
	which shows a linear dependence between the conditional variance and transfer coefficients. Finally, we note that finite detection efficiency $0<\eta<1$, which is unavoidable in experiments and imposes an additional limitation on QND performance, can be incorporated into Eqs.~\eqref{eq:QNDIdeal} by replacing $C \to C\eta$, as described in Appendix~\ref{app:QND}.

\begin{figure}
		\centering
		\includegraphics[width=\columnwidth]{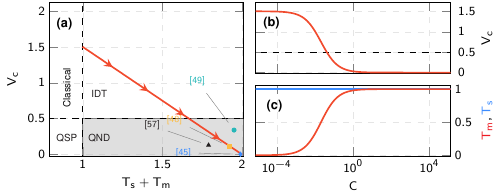}
		\caption{\label{fig:QND}
			$TV$ diagram for ideal QND readout. (a) Conditional variance $V_c$ versus the sum of the transfer coefficients $T_s+T_m$. The points show the position of various experimental implementations of QND measurements in optomechanics and electromechanics with corresponding reference number.
			(b) Conditional variance $V_c$ and (c) transfer coefficients $T_s$ (blue) and $T_m$ (red) as functions of the optomechanical cooperativity $C = 4g^2/\kappa\gamma$ for the same mechanical state. The parameters are the same as in previous figures.}
	\end{figure}

	We plot these quantities in Fig.~\ref{fig:QND}. In panel (a), the $TV$ diagram shows the linear dependence of the conditional variance $V_c$ on the sum of the transfer coefficients $T_s+T_m$ following Eq.~\eqref{eq:QND_Vc}. In the limit $C\to 0$, the conditional variance reduces to the initial mechanical variance $V_x$ and transmission of the signal to the meter vanishes, $T_m\to 0$ ($T_s=1$ independent of cooperativity). Increasing the cooperativity pushes the system through the upper right quadrant corresponding to the IDT regime until the cooperativity reaches the bound~\eqref{eq:C_bound}. The system then crosses to the lower right quadrant of QND measurement. Finally, in the limit $C\to\infty$, we have $V_c\to 0$ and $T_s+T_m\to 2$, corresponding to perfect QND measurement.
	This behavior is further illustrated in Fig. \ref{fig:QND}(b) and (c) where the conditional variance $V_c$ and transfer coefficients $T_{s,m}$ are plotted against the optomechanical cooperativity $C$. Analogous behavior can be observed for any initial mechanical variance $V_x$; for sub-shot noise variance, $V_x\leq\frac{1}{2}$, the system starts below the horizontal line in the $TV$ diagram for $C= 0$ and is thus squarely in the QND regime for any nonzero cooperativity.
	
	The individual data points in Fig.~\ref{fig:QND}(a) show the position of various experimental implementations of QND measurements in optomechanics and electromechanics within the $TV$ diagram. For electromechanical experiments, the cavity input exhibits thermal noise and thus requires modifying the correlation functions~\eqref{eq:cav_in}, which increases the conditional variance $V_c$ and reduces the transfer coefficient $T_m$ as we describe in detail in Appendix~\ref{app:QND}. In the large-temperature limit, $n_c\to\infty$, the conditional variance reduces to the initial mechanical variance, $V_c\to V_x$, and the transfer coefficient disappears, $T_m\to 0$, since the thermal cavity noise overwhelms the detector. However, for all electromechanical experiments included in Fig.~\ref{fig:QND}(a), the thermal noise is only weak and all implementations thus remain in the QND regime.

	\section{Imperfect quantum nondemolition measurements}\label{sec:imperfections}

	\subsection{Overview of experimental imperfections}
	
	To go beyond the paradigm of ideal QND interaction, we now consider several extensions of the QND Hamiltonian $H_{\rm QND}$. These extensions can be collected in the total Hamiltonian
	\begin{align}\label{eq:HFull}
		\begin{split}
			H &= \frac{\delta_c}{2}(X^2+Y^2) + \mu x^2+\nu p^2 + \frac{\xi}{2}(xp+px)  \\
			&\quad + 2gXx+ 2gX[x\cos(2\omega_m t)+p\sin(2\omega_m t)]
		\end{split}
	\end{align}
	with free oscillations of the cavity field at frequency $\delta_c$, quadratic evolution of the mechanical quadratures at rates $\mu,\nu$, mechanical position squeezing with rate $\xi$, and a fast oscillating term accounting for extension of the QND Hamiltonian beyond the rotating wave approximation (RWA). The second and third term can, alternatively, be expressed in terms of free oscillations at frequency $\delta_m$ and squeezing at rate $\zeta$, $\mu x^2+\nu p^2 = \frac{1}{2}\delta_m(x^2+p^2)+\frac{1}{2}\zeta(x^2-p^2)$, where $\delta_m = \mu+\nu,\zeta=\mu-\nu$, and lead to mixing of the position and momentum quadratures, degrading the quality of the QND measurement.
	
	In practice, free oscillations of both modes can arise through a detuning of the mechanical oscillator, cavity resonance, and the sideband drives. If the cavity mode is driven with pumps at general frequencies $\omega_\pm$ which differ from the ideal mechanical sidebands $\omega_c\pm\omega_m$, the interaction is described by the Hamiltonian
	\begin{equation}
		H = \frac{\delta_c}{2}(X^2+Y^2) + \frac{\delta_m}{2}(x^2+p^2) + 2gXx,
	\end{equation}
	where $\delta_c = \omega_c - \frac{1}{2}(\omega_++\omega_-)$, $\delta_m =\omega_m - \frac{1}{2}(\omega_+-\omega_-)$ and we assume $\delta_{c,m}\ll\kappa$ so we can apply the rotating wave approximation. Parametric amplification of the mechanical motion---described by the terms $\frac{1}{2}\zeta(x^2-p^2)+\frac{1}{2}\xi(xp+px)$---is possible for devices that exhibit non-negligible Duffing nonlinearity~\cite{Setter2019,Huber2020}
	or for coherent scattering of pairs of photons between the mechanical sidebands in levitodynamics~\cite{Cernotik2020}. Finally, the terms oscillating at $2\omega_m$ become relevant for finite sideband ratio $\kappa/\omega_m$ for which the rotating wave approximation has only a limited validity.
	
	In the following, we describe and analyze the role each term plays in optomechanical QND measurements; although we include all terms in the full Hamiltonian~\eqref{eq:HFull}, we will study each term and its contributions separately for clarity. In particular, we show that (i) free cavity oscillations only reduce the overall measurement efficiency, (ii) of all bilinear mechanical terms, only the $\nu p^2$ term has a detrimental effect on the quality of the QND readout. However, since it can arise through a detuning of the sideband drives or mechanical parametric amplification, it can always be reduced by suitably shifting the driving frequencies. We also show that for moderate sideband ratios, $\kappa\lesssim\omega_m$, optimization of the optomechanical cooperativity allows near-ideal QND measurements for a broad range of system parameters (including detection losses and thermal mechanical noise). Ideal QND measurements of mechanical motion are thus possible in a range of realistic optomechanical devices.

	\subsection{Slow cavity oscillation}\label{ssec:delta}
	
	We start by considering the effect of cavity detuning as described by the Hamiltonian
	\begin{equation}\label{eq:QND_dc}
		H = \delta_c a^\dagger a + g(a+a^\dagger)(b+b^\dagger).
	\end{equation}
	For this Hamiltonian and the standard dissipation of both modes, the conditional variance and transfer coefficients are straightforward to calculate from the scattering matrix as we detail in Appendix~\ref{app:imperfections}. Although the cavity oscillations mix the amplitude and phase quadrature (and therefore, seemingly, the imprecision and backaction noise), the measurement remains QND with efficiency reduced according to
	\begin{equation}\label{eq:C_delta_c}
		C\to \frac{\kappa^4}{(\kappa^2+4\delta_c^2)^2}C.
	\end{equation}
	This apparent contradiction can be explained when noting that the cavity field still interacts with a single mechanical quadrature and all measurement backaction accumulates in the complementary quadrature. The only result of the cavity oscillations is to transfer part of the information about the measured position quadrature of the mechanical mode into the unmeasured amplitude quadrature of the output field, which manifests as pure loss. This loss can be quantified by the competition between information transfer to the unmeasured quadrature (at rate $\delta_c$) and leakage out of the cavity (rate $\kappa$), resulting in the prefactor on the right-hand side of Eq.~\eqref{eq:C_delta_c}.

	\subsection{Bilinear mechanical dynamics}\label{ssec:squeezing}
	
	Next, we analyze the effects of additional mechanical evolution during the QND measurement as described by the Hamiltonian
	\begin{equation}
		H = \mu x^2+\nu p^2 + \frac{\xi}{2}(xp+px) + 2gXx.
	\end{equation}
	The quadratic position term $\mu x^2$ does not affect the quality of the QND readout since it only affects the equation of motion for the unmeasured momentum quadrature,
	\begin{equation}
		\dot{p} = -2\mu x - 2gX - \frac{\gamma}{2}p + \sqrt{\gamma}p\inpt.
	\end{equation}
	This term thus introduces additional noise in the mechanical momentum quadrature but, because this quadrature is decoupled from the mechanical position quadrature and the optical phase quadrature, it does not reduce the quality of the QND readout. On the other hand, the quadratic momentum term $\nu p^2$ introduces a similar term in the equation of motion for the position quadrature,
	\begin{equation}\label{eq:EOM_x}
		\dot{x} = 2\nu p - \frac{\gamma}{2}x + \sqrt{\gamma}x\inpt,
	\end{equation}
	and therefore feeds the measurement backaction to the output phase quadrature, destroying the QND nature of the measurement.
	
	\begin{figure}
		\centering
		\includegraphics[width=\linewidth]{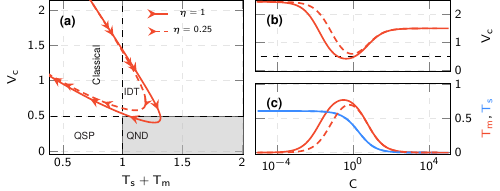}
		\caption{\label{fig:Imp}
			$TV$ diagram for imperfect QND readout. (a) Conditional variance $V_c$ versus the sum of transfer coefficients $T_s+T_m$ for unit detection efficiency ($\eta = 1$, solid) and limited detection efficiency ($\eta = 0.25$, dashed). (b) Conditional variance $V_c$ and (c) transfer coefficients $T_s$ (blue) and $T_m$ (red) versus optomechanical cooperativity. The initial mechanical state is the same as before; in addition, we have $\nu/\gamma = 0.1$, $\mu=\xi=0$.}
	\end{figure}
	
	We study the effects of the quadratic momentum term $\nu p^2$ on the quality of QND readout numerically in Fig.~\ref{fig:Imp} for $\nu/\gamma = 0.1$. Due to the measurement backaction contaminating the optical output, the readout starts in the fully classical regime where $T_s+T_m<1$, $V_c>\frac{1}{2}$. As the cooperativity is increased, the readout improves and reaches, through the IDT region, the regime of QND readout. Further enhancement of the cooperativity beyond unity degrades the measurement (via increasing the strength of measurement backaction) which then leaves the QND regime. While the qualitative characteristics remain valid for any strength of the quadratic momentum term $\nu$, the quantitative details---particularly the possibility of reaching the QND regime---generally depend on the strength of these free dynamics and detection efficiency; unlike in the ideal QND regime, detection losses do not simply rescale the cooperativity as can be seen from the dashed line in Fig. \ref{fig:Imp}, which assumes detection efficiency $\eta = 0.25$. The minimum conditional variance attainable for given $\nu$ by optimizing over the optomechanical cooperativity allows us to define a \emph{generalized} SQL, which we further analyze in the following. 
	
\begin{figure}
		\centering
		\includegraphics[width=\linewidth]{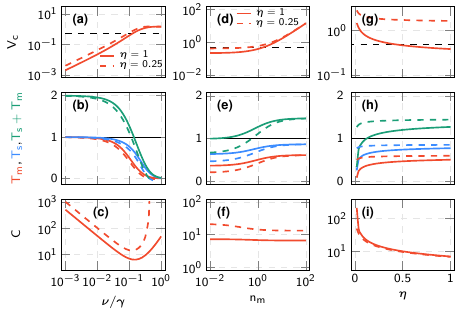}
		\caption{\label{fig:SQL}
			Generalized standard quantum limit for imperfect QND readout. (a) Minimum conditional variance $V_c$, (b) corresponding transfer coefficients $T_s$ (blue), $T_m$ (red), and their sum $T_s+T_m$ (green), and (c) optimal optomechanical cooperativity versus the strength of the quadratic momentum term $\nu$ for $n_m=1$, $m_{\rm sq}=0$. The solid lines show results for perfect detection, $\eta = 1$, dashed lines are for $\eta = 0.25$. 
			(d)--(f) Conditional variance, transfer coefficients, and optimal cooperativity plotted against thermal mechanical noise $n_m$ for $\nu/\gamma=0.1$.
			(g)--(i) The same parameters as functions of the detection efficiency $\eta$ for $\nu/\gamma = 0.1$, $n_m = 1$ (solid lines) and $n_m=10$ (dashed).
			In all plots, $\mu=\xi = 0$; all other parameters are the same as in the previous figures.}
	\end{figure}

	We analyze the generalized SQL in detail in Fig.~\ref{fig:SQL} where we plot (a) the minimum conditional variance, (b) transfer coefficients, and (c) optimal cooperativity as functions of the strength of the quadratic momentum term $\nu$. For $\nu\to 0$, we recover the ideal QND readout with $V_c\to 0$ and $T_{s,m}\to 1$ for cooperativity $C\to\infty$. As the strength of the momentum term increases, the conditional variance grows as well, eventually surpassing the shot-noise level for $\nu/\gamma = 0.125$ (for perfect detection with $\eta = 1$; for limited detection efficiency with $\eta=0.25$, the bound is reduced to $\nu/\gamma = 0.089$). At the same time, the transfer coefficients gradually decrease: For perfect detection ($\eta=1$), their sum drops below unity for $\nu/\gamma = 0.14$ ($\nu/\gamma = 0.115$ for $\eta=0.25$).
	
	In Fig. \ref{fig:SQL}(d)--(f), we study the influence of the initial mechanical state (especially the mean occupation of the mechanical bath $n_m$) on the possibility to reach the QND regime.
	While the optimal cooperativity does not change dramatically with the initial noise of the mechanical mode, this noise strongly influences the attainable conditional variance in the presence of the quadratic momentum term $\nu$. The noise level at which the measurement is no longer QND generally depends on the strength of the squeezing term with stronger squeezing being more sensitive to thermal noise.
	For $\nu/\gamma = 0.1$ (used in Fig.~\ref{fig:SQL}), the conditional variance crosses the boundary for $n_m = 1.81$ (assuming perfect detection, $\eta = 1$); limited detection efficiency of $\eta = 0.25$ reduces this value to $n_m = 0.49$.
	Initial correlations between the position and momentum quadratures (characterized by $\Im{}m\neq 0$) lead to a modification of the conditional variance and transfer coefficients, making it possible to reach $T_{s,m}>1$ \cite{Roch1992} (see also appendix~\ref{app:imperfections}); these correlations, however, have no role in the ideal QND readout.
	
	Finally, we analyze the generalized SQL as a function of the detection efficiency in Fig.~\ref{fig:SQL}(g)--(i). Again, we can see that the conditional variance is more sensitive than the transfer coefficients, i.e., there is a region of detection efficiencies where $V_c>\frac{1}{2}$ but $T_s+T_m > 1$, which takes the measurement out of the optimal QND regime and into the region of the information--damage trade-off. The optimal cooperativity remains of the order of 10 for most values of efficiency and increases sharply for $\eta\ll 1$, where it is needed to compensate for the fast loss of signal in the optical output.
	
	Since the quadratic position and momentum terms $\mu x^2+\nu p^2$ are most likely to appear through parametric amplification and residual free oscillations of the mechanical motion and since only the quadratic momentum term degrades the quality of the QND measurement, these effects can be used to compensate one another and reach ideal optomechanical QND measurement. Crucially, if the mechanical mode exhibits squeezing $\frac{1}{2}\zeta(x^2-p^2)$ (caused by Duffing nonlinearity or coherent scattering between sidebands in levitodynamics), free mechanical oscillations $\frac{1}{2}\delta_m(x^2+p^2)$ can compensate its effects on QND measurement when $\delta_m = \zeta$, which cancels the quadratic momentum term. Since the oscillation frequency $\delta_m$ is set by the detuning of the driving tones from the mechanical sidebands, it is fully under experimental control and can be adjusted as needed.
	
	The final squeezing term $\frac{1}{2}\xi(xp+px)$ does not change the nature of the QND readout but its effect has to be included in the processing of the measurement data and interpreting the results. This term does not mix the two mechanical quadratures but instead amplifies (or squeezes) the position quadrature $x$ as can be seen from the equations of motion for mechanical quadratures,
	\begin{subequations}
		\begin{align}
			\dot{x} &= \xi x-\frac{\gamma}{2}x+\sqrt{\gamma}x\inpt,\\
			\dot{p} &= -2gX-\xi p-\frac{\gamma}{2} + \sqrt{\gamma}p\inpt.
		\end{align}
	\end{subequations}
	This squeezing of the mechanical position quadrature naturally modifies the attainable conditional variance (and, to some extent, the transfer coefficients; see Appendix~\ref{app:imperfections}) but it is necessary that its contribution be evaluated separately to allow precise characterization of the measurement itself. Strong squeezing can result in $V_c<\frac{1}{2}$ for cooperativity that does not allow truly QND measurement without this squeezing; on the other hand, mechanical amplification might make it seem that the QND regime is unattainable due to large conditional variance, $V_c>\frac{1}{2}$, but this can be caused by the large antisqueezing of the mechanical position quadrature.
	
	This consequence of mechanical amplification can be seen more formally from the conditional variance and transfer coefficients,
	\begin{subequations}\label{eq:VTT_xi}
		\begin{align}
			V_c &= \frac{1}{V_\xi^{-1}+32C\gamma^2/(\gamma+2\xi)^2}, \\
			T_s &= 1, \\
			T_m &= \frac{32C\gamma^2/(\gamma+2\xi)^2}{V_\xi^{-1}+32C\gamma^2/(\gamma+2\xi)^2}.
		\end{align}
	\end{subequations}
	The effect of parametric amplification is therefore twofold: First, it changes the variance of the mechanical resonator from $V_x = n_m+\Re m_{\rm sq}+\frac{1}{2}$ to $V_\xi = V_x(\gamma+2\xi)^2/(\gamma-2\xi)^2$, which results in apparent reduction (for $\xi<0$) or increase ($\xi>0$) of the mechanical variance. Second, it modifies the measurement efficiency according to $C\to C\gamma^2/(\gamma+2\xi)^2$, which leads to signal amplification for $-\gamma < \xi<0$. Other than these modifications, the measurement remains QND.

	\subsection{Counterrotating terms}\label{ssec:Floquet}
	
	\begin{figure}
		\centering
		\includegraphics[width=\linewidth]{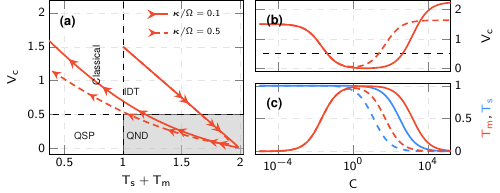}
		\caption{\label{fig:beyondRWA}
			QND readout with counterrotating terms. (a) Conditional variance $V_c$ plotted against the sum of transfer coefficients $T_s+T_m$ for a variable cooperativity $C$ and a fixed sideband ratio $\kappa/\omega_m = 0.1$ (solid) and $\kappa/\omega_m = 0.5$ (dashed). (b) Conditional variance $V_c$ and (c) transfer coefficients $T_{s,m}$ versus optomechanical cooperativity for the data in (a). The system parameters are $\gamma/\omega_m = 0.01$, $n_m = 1$, $m_{\rm sq}= 0$, and detection efficiency $\eta = 1$.}
	\end{figure}
	
	The ideal QND Hamiltonian $H_{\rm QND}$ is obtained from the optomechanical interaction under the rotating wave approximation which is valid only in the resolved sideband regime, $\kappa\ll\omega_m$. In general, finite sideband resolution leads to mixing of the mechanical quadratures and reduces the quality of the QND readout. To see and quantify its impact, we start from the QND interaction with the counterrotating terms included,
	\begin{equation}\label{eq:H_QND_full}
		H = 2gXx + 2gX[x\cos(2\omega_m t)+p\sin(2\omega_m t)].
	\end{equation}
	The effect of the counterrotating terms oscillating at $2\omega_m$ on the dynamics can be treated perturbatively using Floquet techniques~\cite{Malz2016,Malz2016a}: We start by expressing the time-dependent drift matrix corresponding to the Hamiltonian~\eqref{eq:H_QND_full} in terms of its frequency components (see also Appendix~\ref{app:Floquet}),
	\begin{equation}
		{\bf A}(t) = {\bf A}^{(-1)} e^{-2i\omega_m t} + {\bf A}^{(0)} + {\bf A}^{(1)} e^{2i\omega_m t},
	\end{equation}
	where the matrices ${\bf A}^{(j)}$ are time independent. We can express the vector of quadrature operators in the same manner (albeit with an infinite expansion),
	\begin{equation}
		{\bf u}(t) = \sum_{n = -\infty}^\infty {\bf u}^{(n)}(t) e^{2in\omega_m t},
	\end{equation}
	which allows us to express the equations of motion in terms of the individual frequency components ${\bf u}^{(n)}(t)$ which are coupled to ${\bf u}^{(n\pm 1)}(t)$ via ${\bf A}^{(\mp 1)}$. We can then solve the equations of motion for ${\bf u}^{(n)}$ in frequency space and obtain the output field by a summation of the Floquet components at the corresponding frequencies,
	\begin{equation}
		{\bf u}_{\rm out}(\omega) = \sum_n {\bf u}_{\rm out}^{(n)}(\omega+2n\omega_m).
	\end{equation}
	For moderate sideband ratio, $\kappa/\omega_m\lesssim 1$, we can truncate the expansion at $n\in\{-1,0,1\}$ and obtain a modified scattering matrix, conditional variance, and transfer coefficients as detailed in Appendix~\ref{app:Floquet}.
	
	The resulting $TV$ diagram is plotted in Fig.~\ref{fig:beyondRWA} for sideband ratio $\kappa/\omega_m = 0.1$ (solid) and $\kappa/\omega_m = 0.5$ (dashed). At first, the measurement follows the ideal QND readout (cf. Fig.~\ref{fig:QND}), crossing from the IDT regime over to the QND regime and close to the perfect QND measurement ($V_c\approx 0$, $T_s+T_m\approx 2$ for $C\sim 1$). As the cooperativity is further increased, the measurement starts to deteriorate (especially the transfer coefficients) and moves through the QSP regime to the classical regime due to increased measurement backaction. This general behavior can be seen for both values of sideband ratio shown in Fig.~\ref{fig:beyondRWA}, demonstrating that realistic optomechanical devices with finite sideband ratio can reach deep into the QND regime with optimized optomechanical cooperativity of the order of unity.
	
	\begin{figure}
		\includegraphics[width=\linewidth]{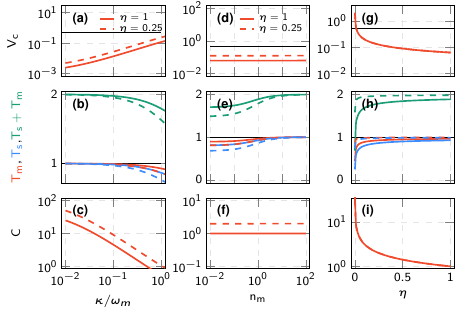}
		\caption{\label{fig:gSQL_RWA}
			Generalized SQL for QND readout beyond RWA. (a) Minimum	 conditional variance $V_c$, (b) corresponding transfer coefficients $T_s $ (blue), $T_m$ (red), and their sum $T_s + T_m$ (green), and (c) optimal optomechanical cooperativity versus the sideband ratio $\kappa/\omega_m$ for $\gamma/\omega_m = 0.01$, $n_m = 1$, $m_{\rm sq}= 0$. Solid lines show results for perfect detection, $\eta = 1$, dashed lines are for $\eta = 0.25$. (d)--(f) Conditional variance, transfer coefficients, and optimal cooperativity plotted against the thermal occupation of the mechanical mode $n_m$ for the sideband ratio $\kappa/\omega_m = 0.5$. (g)--(i) The same quantities plotted versus detection efficiency $\eta$ for $\kappa/\omega_m = 0.5$, $n_m = 1$ (solid lines) and $n_m=10$ (dashed).
		}
	\end{figure}
	
	We further investigate the generalized SQL in Fig.~\ref{fig:gSQL_RWA} which shows that the QND regime can be reached for a broad range of experimental parameters. Crucially, the conditional variance remains below the shot-noise level for sideband ratios $\kappa/\omega_m\leq 1$; note that this result does not rely on the truncation of the Floquet basis to $n\in\{-1,0,1\}$ which we only use to obtain analytical expressions for the covariance matrix and transfer coefficients in Appendix~\ref{app:Floquet}. Moreover, as the sideband ratio is increased, the optimal cooperativity decreases (recall that we recover optimal QND measurement in the limit $\kappa/\omega_m\to 0$ for which the optimum cooperativity diverges), demonstrating that efficient QND measurement with a finite sideband ratio is generally available in suitable experimental regimes. Plotting the generalized SQL against the thermal mechanical noise [Fig.~\ref{fig:gSQL_RWA}(d)--(f)] and detection efficiency [panels (g)--(i)] further proves that the QND measurement is attainable over a broad range of experimental situations.

	\section{Quantum nondemolition readout in levitodynamics}\label{sec:levitation}
	
	\subsection{QND measurement via tweezer modulation}
	
	The deeper understanding of QND measurements and how various imperfections affect their quality allows us to design a suitable strategy for these measurements in levitated optomechanics with the help of coherent scattering. Levitated nanoparticles present an attractive platform for optomechanics due to their thermal isolation and possibility of engineering nonlinear potentials~\cite{Millen2020,Gonzalez-Ballestero2021} but the usual radiation pressure is typically weak due to their small size, requiring strong driving which enhances recoil heating~\cite{Kiesel2013,Jain2016,Delic2020a}. The solution is to use coherent scattering as a coupling mechanism, where the nanoparticle coherently scatters tweezer photons into an empty cavity mode~\cite{Delic2019,Windey2019,Gonzalez_Ballestero2019}. Detuning between the tweezer and cavity mode can be used to switch between beam splitter (which has been successfully used to cool nanoparticle motion to its quantum ground state~\cite{Delic2020,Pontin2023}) and two-mode squeezing (crucial for generating squeezing~\cite{Cernotik2020} and entanglement~\cite{Chauhan2020,Rudolph2020,Brandao2021,Chauhan2022}) interaction in analogy with detuning of the driving laser. The resulting interaction can reach the strong~\cite{delosRiosSommer2021,Ranfagni2021} and even ultrastrong coupling regime~\cite{Kustura2022,Dare2023X}. QND measurements can be engineered using a suitably modulated tweezer as we describe in detail below. This setup is analogous to two-tone driving in standard cavity optomechanics with additional mechanical squeezing due to coherent scattering between the two sidebands.
	
	The interaction of nanoparticle motion with an optical field is given by
	\begin{equation}\label{interaction_Hamiltonian}
		H_{\rm int} = -\frac{1}{2}\alpha_p {\bf E}^2(x_m,t)
	\end{equation}
	where  $\alpha_p$ is the nanoparticle polarizability and ${\bf E}(x_m,t)$ is the electric field at the particle position $x_m$ (not to be confused with the mechanical quadrature operator $x$). We can then decompose the electric field into the tweezer and cavity components, ${\bf E} =   {\bf E}_{\rm tw} +  {\bf E}_{\rm cav}$, and substitute this expansion back into Eq.~\eqref{interaction_Hamiltonian}, obtaining three terms (see Ref.~\cite{Gonzalez_Ballestero2019} and Appendix~\ref{app:levitation} for details). The terms proportional to ${\bf E}_{\rm tw}^2$ and ${\bf E}_{\rm cav}^2$ give rise, respectively, to the trapping potential and radiation pressure optomechanical interaction; the latter can be neglected since it is weak and can be avoided altogether by suitably positioning the nanoparticle within the cavity standing wave. The cross term ${\bf E}_{\rm tw}\cdot{\bf E}_{\rm cav}$ describes coherent scattering of tweezer photons into the cavity field. The total Hamiltonian (cavity field, optical trapping, and coherent scattering) is then given by
	\begin{equation}\label{eq:Hlev}
		H = \Delta a^\dagger a + \frac{p_m^2}{2m} + \frac{1}{2}m\omega_{\rm tr}^2x_m^2  - G(a+a^\dagger)x_m,
	\end{equation}
	where $\Delta = \omega_{\rm cav} - \omega_{\rm tw}$ is the detuning between the cavity and tweezer frequencies, $p_m$ is the nanoparticle momentum, and $m$ its mass. The precise values of the trapping frequency $\omega_{\rm tr}$ and coupling rate $G$ depend on the properties of the nanoparticle, tweezer, and cavity mode.
	
	To implement a QND measurement of the nanoparticle motion, the tweezer amplitude $\mathbf{E}_{\rm tw}$ can be modulated at a frequency $\Omega$. This modulation results in the trapping frequency and optomechanical interaction being time dependent~\cite{Cernotik2020} (see also Appendix~\ref{app:levitation}),
	\begin{subequations}\label{eq:modulation}
		\begin{align}
			\frac{1}{2}m\omega_{\rm tr}^2x_m^2 &\to \frac{1}{2}m\omega_{\rm tr}^2[1+\alpha\cos(\Omega t+\phi)]^2x_m^2, \\
			G(a+a^\dagger)x_m &\to G[1+\alpha\cos(\Omega t+\phi)](a+a^\dagger)x_m,
		\end{align}
	\end{subequations}
	where $\alpha$ and $\phi$ are the modulation depth and phase, respectively. We can now introduce the mechanical annihilation operator
	\begin{equation}
		b = \sqrt{\frac{m\omega_m}{2}}\left(x_m+\frac{i}{m\omega_m}p_m\right)
	\end{equation}
	with the mechanical frequency related to the trapping frequency via
	\begin{equation}
		\omega_m = \omega_{\rm tr}\sqrt{1+\frac{1}{2}\alpha^2};
	\end{equation}
	next, we set the tweezer on resonance with the cavity, $\omega_{\rm tw} = \omega_{\rm cav}$, move to a frame rotating with the modulation frequency $\Omega$ for the mechanical mode, and apply the rotating wave approximation (valid in the resolved-sideband regime, $\kappa<\omega_m$) to obtain the interaction Hamiltonian
	\begin{equation}\label{eq:HQNDlev}
		H = \frac{\omega_m-\Omega}{2}(x^2+p^2) + \frac{\alpha^2\omega_m}{16(2+\alpha^2)}(x^2-p^2) - \frac{1}{2}{\alpha g} Xx,
	\end{equation}
	where we returned back to the mechanical quadrature operators $x = (b+b^\dagger)/\sqrt{2}$, $p = -i(b-b^\dagger)/\sqrt{2}$. We also introduced $g = G/\sqrt{2m\omega_m}$ and set $\phi = 0$ without loss of generality. To achieve QND readout without unwanted backaction in the position quadrature $x$, we set the modulation frequency as
	\begin{equation}
		\Omega = \omega_m\frac{16 + 7\alpha^2}{16+8\alpha^2}
	\end{equation}
	which allows us to eliminate the momentum degree of freedom from Eq.~(\ref{eq:HQNDlev}) and get the final Hamiltonian for QND measurement in levitodynamics
	\begin{equation}\label{eq:QNDlev}
		H_{\rm QND} = \frac{\alpha^2\omega_m}{8(2+\alpha^2)}x^2 - \frac{1}{2}{\alpha g} Xx,
	\end{equation}
	as we discussed previously, the first term introduces backaction only in the unmeasured momentum quadrature and therefore does not affect the quality of the QND measurement which is the result of the second term.
	
	State of the art levitodynamic systems reach quantum cooperativity of the order of unity and above~\cite{Delic2020,delosRiosSommer2021,Dare2023X}, allowing efficient QND measurements as well. Recall from Sec.~\ref{ssec:QND} that we need $C>\frac{1}{16}$ for single-quadrature measurement to reach the QND regime. From Eq.~\eqref{eq:QNDlev}, we have the QND coupling rate $\alpha g/2$, which reduces the cooperativity (compared with unmodulated interaction) by a factor of $\frac{1}{4}\alpha^2$. Moderate modulation of the tweezer, $\alpha\in (0.01,0.1)$, can therefore be expected to allow QND measurements in these systems. However, a direct quantitative estimation is difficult due to different nature of decoherence processes compared with other optomechanical systems. In standard optomechanical systems, decoherence is caused by energy exchange with the (solid-state) environment modelled as an exchange interaction and resulting in the thermal phase-insensitive Lindbladians $\gamma({n_m}+1)\mathcal{D}[b]\rho+\gamma{n_m}\mathcal{D}[b^\dagger]\rho$. On the other hand, levitated nanoparticles suffer primarily from gas damping and photon recoil described by a diffusion phase-sensitive term $\Gamma\mathcal{D}[x]\rho$~\cite{Gonzalez_Ballestero2019}, which are stronger than in other optomechanical systems due to the small mass of nanoparticles, strong fields required for trapping, and lack of clamping that would allow heat to be efficiently dissipated. This amplitude diffusion leads to dynamical instability and therefore requires active stabilization, resulting in an effective thermal decoherence rate, which can then be used to evaluate the quantum cooperativity. Nevertheless, we describe all mechanical dissipation using an effective thermal-noise model which is equivalent to the localization noise in the limit of large temperature and mechanical frequency, $\Gamma\mathcal{D}[x]\rho\simeq \Gamma(\mathcal{D}[b]\rho+\mathcal{D}[b^\dagger]\rho)\simeq \gamma_{\rm eff}({n}_{\rm eff}+1)\mathcal{D}[b]\rho+\gamma_{\rm eff}{n}_{\rm eff}\mathcal{D}[b^\dagger]\rho$.
	In addition, fluctuations of the trapping light intensity and phase give rise to effective trap displacement (position localization) noise, which can be described within the same master-equation framework as an additional diffusion channel acting on the particle position~\cite{Gonzalez_Ballestero2019}. This contribution originates from stochastic fluctuations of the optical potential minimum and is formally equivalent to an extra position-diffusion term in the Langevin description.
	
	We also note that the tweezer modulation results in modulated photon recoil heating of the nanoparticle motion and might thus affect the observed decoherence mechanisms. However, within the Born-Markov approximation, the nanoparticle interacts with bath modes at the mechanical frequency $\omega_m$ and sidebands $\omega_m \pm \Omega$. By assuming a flat bath density of states, these sideband interactions average to yield a time-independent, renormalized recoil heating rate $\Gamma_{\rm rec} = \Gamma_{\rm rec}^0(1 + \alpha^2/2)$. The modulation results only in a perturbative increase to the heating rate in absence of modulation $\Gamma_{\rm rec}^0$ rather than a time-dependent dissipation rate. Therefore, the QND character of our protocol remains fully preserved. Since photon recoil is only a part of the total decoherence of the nanoparticle and the renormalization term generally small, we do not model this effect in further calculations.
	
	One final complication stems from the fact that the total trapping intensity is limited to keep absorption heating of the particle sufficiently low and prevent its melting. Simultaneous application of multiple tweezers---typically one for quantum state preparation and one for state measurement---is then subject to a trade-off between efficient state preparation and measurement. This problem can be resolved by performing state preparation and measurement sequentially but this results in reheating of the particle motion during the measurement. In the following, we describe both approaches and outline the trade-off they offer between the meter transfer coefficient $T_m$ on one hand and the system transfer coefficient $T_s$ and conditional variance $V_c$ on the other; the main difference lies in the choice of pulse shape function used for filtering the signal.

	\subsection{State tomography in two-tweezer setups}

\begin{figure}
			\includegraphics[width=\linewidth]{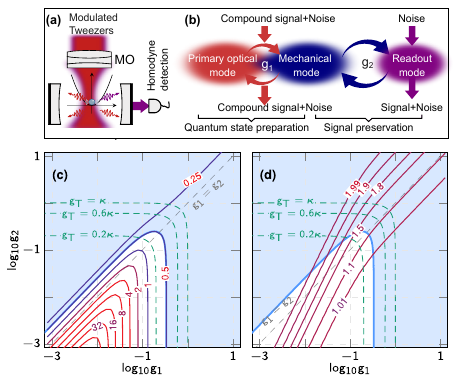}
			\caption{\label{fig:DualTweezer}
				QND readout in dual-tweezer levitodynamics. (a) Schematic of the proposed experimental setup for quantum state preparation and measurement of a levitated nanoparticle via coherent scattering with two tweezers and two cavity modes and (b) depiction of the interactions involved. One of the two trapping beams (red) is used to dissipatively stabilize a quantum state of interest, while the other (purple) is used for its measurement. (c) Conditional variance $V_c$ and (d) the sum of transfer coefficients $T_s+T_m$ for dissipative squeezing experiment with $\alpha_1=\alpha_2=0.2$, $V_x=10^7$, $\gamma/\kappa_1=10^{-9}$, and $\kappa_1=\kappa_2$. The blue region corresponds to the QND regime of the measurement. The dashed green lines correspond to fixing the total trapping intensity, corresponding to constant ${g_1^2+g_2^2} = g_T^2$.}
		\end{figure}

		First, we consider QND readout applied simultaneously with another interaction used for state preparation. The simplest such examples are measuring the mechanical state in a cooling or dissipative squeezing experiment. We now expand the Hamiltonian derived in the previous section to include two tweezer beams coherently scattering photons into two distinct cavity modes as shown in Fig.~\ref{fig:DualTweezer}(a). The primary tweezer (red) cools the mechanical motion; when suitably modulated, it is capable of generating a squeezed mechanical state~\cite{Cernotik2020}. The (generally weaker) readout tweezer (purple) is used for probing the quantum mechanical state generated by the primary tweezer. Since the total trapping intensity, which sets the frequency of particle motion, is limited by absorption heating of the particle material, the key question concerns the optimal distribution of intensity between the two tweezers. Strong primary tweezer results in efficient state preparation but inefficient measurement; the readout can be improved by increasing the readout tweezer intensity but this reduces the fidelity of state preparation since the intensity of the primary tweezer has to be appropriately reduced.
		
		To derive the total system Hamiltonian, we assume that the two tweezers and cavity modes they scatter into are sufficiently far away in frequency so there can be no scattering between the primary and readout subsystem; this is a reasonable assumption since the free spectral range (several gigahertz for Fabry--P\'{e}rot resonators a few centimeters long) is typically much larger than the mechanical frequency (typically below 100 kilohertz). We thus obtain the total Hamiltonian by adding the Hamiltonians corresponding to the two scattering processes (see Appendix~\ref{app:dual_tweezer} for details),
		\begin{align}\label{eq:dual_tweezer}
			\begin{split}
				H &= \frac{{{\omega _m} - {\Omega _2}}}{2}({x^2} + {p^2}) + {{\tilde \alpha }_1}({x^2} - {p^2})\\
				&\quad- \frac{{{g_1}}}{2}({X_1}x + {Y_1}p) - \frac{{{g_1}{\alpha _1}}}{4}({X_1}x - {Y_1}p)\\
				&\quad+ {{\tilde \alpha }_2}\left[ {({x^2} - {p^2})\cos 2\phi  - (xp + px)\sin 2\phi } \right]\\
				&\quad- \frac{{{g_2}{\alpha _2}}}{2}{X_2}(x\cos \phi  - p\sin \phi ),
			\end{split}
		\end{align}
		where $\tilde{\alpha}_1 = {{{\alpha _1}\omega _{1,{\rm{tr}}}^2}}/{{4{\omega _m}}}$ and $\tilde{\alpha}_2 = {{\alpha _2^2\omega _{2,{\rm{tr}}}^2}}/{{16{\omega _m}}}$. This Hamiltonian shows the combined dissipative squeezing dynamics induced by the primary optical tweezer and QND readout due to the second tweezer with the corresponding parametric squeezing processes. In a frame rotating at the modulation frequency of the second tweezer (first term on the right-hand side), the primary tweezer with modulation depth $\alpha_1$ and coupling strength $g_1$ results in parametric squeezing (the second term on the right hand side) and dissipative squeezing (second line); in the limit where the modulation amplitude approaches zero $\alpha_1\to 0$, the Hamiltonian simplifies to the beam-splitter interaction of sideband cooling. The second tweezer with modulation depth $\alpha_2$ and coupling rate $g_2$ allows QND readout of one mechanical quadrature (via the term on the last line) accompanied by an additional parametric squeezing term (third line). Here, we are keeping an arbitrary mechanical quadrature $x_\phi = x\cos\phi-p\sin\phi$ to allow for full state tomography.
		
		An important caveat is that the modulation frequency $\Omega_2$ will depend on the quadrature we are measuring, becoming $\phi$-dependent. This can be seen from the need to remove any terms containing $p_\phi^2 = (x\sin\phi+p\cos\phi)^2$ from the Hamiltonian, as these would introduce measurement backaction noise into the measured quadrature $x_\phi$ since $[x_\phi,p_\phi^2] = 2ip_\phi$. In the following, we will only consider a measurement of the position quadrature $x$ as this quadrature is squeezed by the primary tweezer; we therefore set $\phi = 0$, $\Omega_2 = \omega_m-2\tilde{\alpha}_1-2\tilde{\alpha}_2$, which results in the simplified Hamiltonian (note the absence of any quadratic momentum terms)
		\begin{equation}\label{eq:HDual}
			\begin{split}
				H &= 2({{\tilde \alpha }_1} + {{\tilde \alpha }_2}){x^2} - \frac{{{g_1}}}{2}({X_1}x + {Y_1}p) \\
				&\quad - \frac{{{g_1}{\alpha _1}}}{4}({X_1}x - {Y_1}p) - \frac{{{g_2}{\alpha _2}}}{2}{X_2}x.
			\end{split}
		\end{equation}
		The analysis can be repeated for a general quadrature $x_\phi$ with suitably modified modulation frequency $\Omega_2$ but this does not contain any new physics not captured here.
		
		In the interaction described by Eq.~\eqref{eq:HDual}, the quantity of interest for measurement is the compound signal arising from the mechanical mode and the first optical mode,
		\begin{subequations}
			\begin{align}
				{{\bar x}_{{\rm{in}}}} &= \frac{1}{{2\sqrt {{\gamma _m}} }}\left[ {{g_1}\left( {{\alpha _1} - 2} \right){\chi _1}\sqrt {{\kappa _1}} {Y_{1,{\text{in}}}} + 2\sqrt \gamma  {x_{{\text{in}}}}} \right],\\
				{{\bar p}_{{\rm{in}}}} &= \frac{1}{{2\sqrt {{\gamma _m}} }}\left[ {{g_1}\left( {{\alpha _1} + 2} \right){\chi _1}\sqrt {{\kappa _1}} {X_{1,{\text{in}}}} + 2\sqrt \gamma  {p_{{\text{in}}}}} \right],
			\end{align}
		\end{subequations}
		which encodes the generated mechanical state; here, we introduced the susceptibility of the primary optical mode, $\chi_i(\omega) = \left(\kappa_i - 2i\omega  \right)^{ - 1}$, and optically broadened mechanical line, $\gamma_m = \gamma + g_1^2\left( 1 - {\alpha _1^2}/{4} \right)/\kappa_1$. The initial variances of the compound signal can be expressed as
		\begin{subequations} \label{eq:compund_signal_variances}
			\begin{align}
				\bar{V}_x &= \frac{\gamma }{{{\gamma _m}}}{V_x} + \frac{{g_1^2{{(2 - {\alpha _1})}^2}}}{{8{\gamma _m}{\kappa _1}}}\,,\\
				\bar{V}_p &= \frac{\gamma }{{{\gamma _m}}}{V_p} + \frac{{g_1^2{{(2 + {\alpha _1})}^2}}}{{8{\gamma _m}{\kappa _1}}}\,;
			\end{align}   
		\end{subequations}
		this indicates that the initial variance of mechanical signal in a cooling experiment ($\alpha_1\to 0$) can be reduced down to $g_1^2/(2 \gamma_m \kappa_1)$. Moreover, modulating the primary optical tweezer ($\alpha_1\neq 0$) induces an asymmetry in the noise profile and may lead to mechanical squeezing below the vacuum noise level. By probing this compound signal through the QND interaction with the second optical mode, we can characterize and analyze the mechanical state generated by the first optical mode.
		
		The conditional variance and transfer coefficients can be readily evaluated from the scattering matrix, as detailed in Appendix~\ref{app:dual_tweezer}, 
		\begin{subequations}
			\begin{align} 
				{V_c} &= \frac{{1 + {C_1}(4 - \alpha _1^2)}}{{\left[ {1 + {C_1}(4 - \alpha _1^2)} \right]{{(V_x^s)}^{ - 1}} + 32 \alpha_2^2 {C_2}}}\,,\\
				T_s &= 1\,,\\
				{T_m} &= \frac{{32 \alpha_2^2 {C_2}}}{{\left[ {1 + {C_1}(4 - \alpha _1^2)} \right]{{(V_x^s)}^{ - 1}} + 32 \alpha_2^2 {C_2}}}\,,
			\end{align}
		\end{subequations}
		where we have defined $C_i=g_i^2/(4\gamma \kappa_i)$ as the rescaled cooperativities. In Fig.~\ref{fig:DualTweezer}(c), (d), we show contour plots illustrating the conditional variance $V_c$ and sum of transfer coefficients $T_s+T_m$ as functions of the two optomechanical coupling strengths. The simulation parameters are similar to the parameters in the experiment on ground-state cooling by coherent scattering \cite{Delic2020} except for slower mechanical decoherence, which is possible with longer optical wavelengths resulting in weaker photon recoil. Notably, $T_s+T_m > 1$ everywhere in Fig.~\ref{fig:DualTweezer}(d) so the QND regime is fully determined only by $V_c < \frac{1}{2}$. This happens at readout cooperativity satisfying
		\begin{equation} 
			{C_2} > \frac{{\left[ {{C_1}(4 - \alpha _1^2) + 1} \right]\left[ {2{V_x} - {C_1}\alpha _1^2\left( {3 - \alpha _1^2} \right) - 1} \right]}}{{16 \alpha_2^2\left[ {2{V_x} + {C_1}{{(2 - \alpha _1^2)}^2}} \right]}} > 0\,.
		\end{equation}
		
		\begin{figure}
			\includegraphics[width=\linewidth]{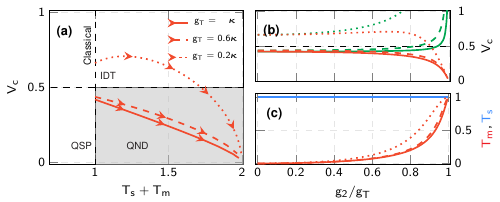}
			\caption{\label{fig:schematic2}
				$TV$ diagram for dual-tweezer levitodynamics. (a) Conditional variance $V_c$ plotted against the sum of transfer coefficients $T_s+T_m$ for different values of $g_T$ when varying the fraction of intensity in the primary/readout tweezer. (b) Conditional variance $V_c$ and (c) transfer coefficients $T_{s,m}$ versus fractional coupling of the readout tweezer corresponding to the data in (a). In panel (b), the green lines show the unconditional squeezing generated by the primary tweezer. The modulation depth for both tweezers $\alpha_1 = \alpha_2 = 0.2$.}
		\end{figure}
		
		As shown in Fig.~\ref{fig:schematic2}, we encounter a fundamental trade-off between quantum state preparation and measurement precision when keeping the sum of the tweezer intensities fixed. Typically, one would expect a weaker intensity for readout to avoid perturbing the quantum state prepared by the primary tweezer. While the measurement has little effect on the mechanical state when keeping $g_2<g_1$ [the conditional mechanical variance is close to the unconditional variance in Fig.~\ref{fig:schematic2}(b)], the measurement might remain close to the QND--IDT boundary for realistic experimental parameters. However, unconditional mechanical squeezing and its QND measurement are possible with state-of-the-art levitated systems---in recent cooling experiments~\cite{Windey2019,Delic2019}, the ratio of the optomechanical coupling rate to the cavity decay rate has reached values as high as $g_i/\kappa_i\sim 0.6$. As our analysis in Fig.~\ref{fig:schematic2} shows, this allows reaching squeezing (both deterministic and conditional) and total system transmission $T_s+T_m\approx1.1$ for $g_1\sim g_2$.
		
		In the opposite limit of strong measurement, $g_2 > g_1$, the conditional variance is reduced, suggesting stronger squeezing than using a combination of parametric and dissipative squeezing (obtained in the limit $g_2\to 0$). Note, however, that this result does not imply stronger squeezing of the mechanical mode as $V_c$ quantifies the variance of the (hypothetical) mechanical output which is related to the localized mechanical mode via the usual input--output relation $b_{\rm out} = \sqrt{\gamma}b-b_{\rm in}$; the precise relation between the conditional variance of the output and localized mode thus depends also on the correlations between the mechanical mode $b$ and its input $b_{\rm in}$. In addition, the conditional variance $V_c$ is evaluated at a specific frequency whereas the total variance of the localized mechanical mode is frequency independent. The results in Fig.~\ref{fig:schematic2} thus do not directly quantify stronger mechanical squeezing attainable via QND measurement than with dissipative and parametric squeezing but rather the ability to measure the mechanical motion with high accuracy.

		\subsection{Utilizing pulsed interaction for QND readout} 
		
		So far, we considered a continuous-driving model to simultaneously prepare and read out the state of a trapped levitated nanoparticle. While this approach benefits from measuring an actively stabilized mechanical state, the readout efficiency (as well as the quality of the prepared state) is limited by the need to have both tweezers (which are subject to a constraint on the total intensity) active at the same time. To alleviate this issue and allow more efficient readout, we now consider a sequential pulsed interaction strategy. This method allows us to achieve a higher efficiency in both control and readout. However, as the particle motion is not actively stabilized during measurement, the signal is partially corrupted by thermalization of the mechanical mode during the QND interaction. We would therefore like to develop tools that allow us to compare the overall performance of this pulsed readout with the dual-tweezer dynamics, and the formalism of the $TV$ diagram is the optimal tool for this task.
		
		First, we prepare the motional mode of the nanoparticle in a specific quantum state by controlling its interaction with the tweezer, taking advantage of coherent scattering into a cavity mode. The dynamics are governed by the Hamiltonian
		\begin{equation}
			H = \frac{{\alpha {\omega _m}}}{{2 + {\alpha ^2}}}{x^2}\, - \frac{g}{2}(Xx + Yp) - \frac{{g\alpha }}{4}(Xx - Yp)\,,
		\end{equation}
		which is the same as Hamiltonian~\eqref{eq:dual_tweezer} without the readout tweezer. This choice ensures that the prepared mechanical state is the same as in the dual-tweezer setup in the limit $g_2\to 0$.
		
		After preparing the mechanical state, the tweezer frequency and modulation is changed to allow QND readout as described by Eq.~\eqref{eq:QNDlev}. At a  time $\tau$ after the start of the measurement, the amplitude quadrature is given by
		\begin{equation}
			x(\tau ) = {e^{ - \frac{1}{2}\gamma \tau }} x(0) + x_B(\tau)\,,
		\end{equation}
		where $x(0)$ is the initial state and $x_B(\tau)$ is the accumulated mechanical noise due to dissipation (see Appendix~\ref{app:pulsed_tweezer} for detailed expression). Note that we assume measurement of the particle's position quadrature $x$ for simplicity; since the measurement is QND, the dynamics of the measured quadrature is always trivial. The scattered light filtered by the cavity provides information about the nanoparticle’s quantum state $x(0)$ depending on the experimental setup and the pulse characteristics. A discrete output mode quadrature ${\cal Y}(\tau)$ can be obtained from the continuous cavity output using a suitable temporal pulse shape function $f_{\rm out}(t)$ via the expression
		\begin{equation}
			{\cal Y}(\tau) = \int_0^\tau ds f_{\rm out}(s)Y_{\rm out}(s)
		\end{equation}
		with the optimal pulse shape determined by the optomechanical dynamics (see Appendix~\ref{app:pulsed_tweezer}). Formal integration then reveals the output phase quadrature to consist of the signal $x(0)$ and various noise terms grouped together in an operator ${\cal Y}_B(\tau)$,
		\begin{align}
			{{\cal Y}_{{\rm{out}}}}(\tau ) &=\sqrt{{\cal G}-1} x(0) +{\cal Y}_B(\tau)\,,
		\end{align}
		where $\cal G$ is the optomechanical measurement gain; We can now use the same methods to calculate the conditional variance and transfer coefficients to obtain the $TV$ diagram for pulsed readout.
		
		\begin{figure}
			\centering
			\includegraphics[width=\linewidth]{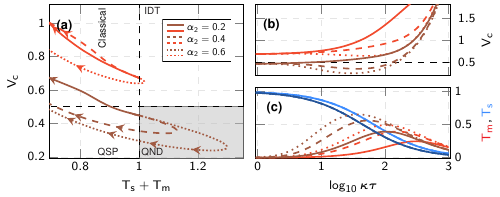}
			\caption{\label{fig:pulsed}
				$TV$ diagram for pulsed QND readout. (a) Conditional variance $V_c$ plotted against the sum of transfer coefficients $T_s+T_m$ with varying measurement pulse duration for different optomechanical couplings. (b) Conditional variance $V_c$ and (c) transfer coefficients $T_{s,m}$ versus pulse duration for the data in (a). The system parameters are $\gamma/\kappa = 10^{-9}$, $n_m = 10^7$. For both state preparation and readout, we use the coupling rates $g = 0.2\kappa$ (light red and blue) and $g = 0.6\kappa$ (dark red and blue); modulation depth $\alpha = 0.2$ is used for state preparation while for readout, we have $\alpha = 0.2$ (solid), $\alpha = 0.4$ (dashed), and $\alpha = 0.6$ (dotted).}
		\end{figure}
		
		The resulting $TV$ diagram is depicted in Fig.~\ref{fig:pulsed}. For state preparation, we assume coupling strength $g = 0.2\kappa$ (light red and blue) and $g = 0.6\kappa$ (dark) with modulation depth $\alpha = 0.2$ for both for consistency with Fig.~\ref{fig:schematic2}. Readout is done using the same coupling strength as state preparation (i.e., $g = 0.2\kappa$ for light red and blue, $g = 0.6\kappa$ for dark) and three different modulation depths, namely $\alpha = 0.2, 0.4, 0.6$ (solid, dashed, and dotted, respectively). The measurement starts with $T_s = 1$, $T_m = 0$ and conditional variance equal to the variance of the initial squeezed state in the limit $\kappa\tau\to 0$ at the border between classical and IDT (for $g = 0.2\kappa$) and between QSP and QND regimes (for $g = 0.6\kappa$). In the opposite limit, $\kappa\tau\to\infty$, we have $T_{s,m}\to 0$ and an increasing variance. In between these limits, the transfer coefficient $T_s$  decreases due to decoherence while the coefficient $T_m$ reaches a maximum at intermediate values of $\kappa\tau$. For sufficiently strong coupling (or modulation depth as their product $g\alpha$ determines the measurement rate), the conditional variance drops as well, reaching the QND regime.
		Notably, our numerical analysis predicts conditional mechanical squeezing of up to $3.4 \,{\rm dB}$. These values are highly competitive with current experimental benchmarks in levitated systems, such as the $4.9 \,{\rm dB}$ squeezing reported in \cite{Kamba2025}, and exceed the $2.7 \,{\rm dB}$ squeezing below thermal motion demonstrated via trapping potential modulation in free space \cite{Rashid2016}.
		
		The limit $\kappa\tau\to 0$ corresponds to the limit $g_2\to 0$ in the dual-tweezer setup. One might expect that the opposite limits (i.e., $\kappa\tau\to\infty$ for pulsed readout and $g_2\to g_T$ for dual-tweezer measurement) coincide as well but this is not the case due to different pulse shape functions. The pulsed readout is optimized for obtaining information about the initial mechanical state and the pulse shape function $f_{\rm out}$ is a decaying exponential; in the dual-tweezer setup, the measurement is optimized for the steady state and is therefore filtered by a constant pulse shape function.
		
		In the intermediate regime, both schemes are capable of reaching the QND regime but with a few important distinctions. The dual-tweezer setup reaches considerable meter transfer coefficient $T_m$ only for strong readout tweezer, $g_2\sim g_T$, which limits the intensity of the primary tweezer. In addition, increasing $g_T$ requires larger ratio $g_2/g_T$ to reach the same value of $T_m$ [see Fig.~\ref{fig:DualTweezer}(c)] which indicates that obtaining strong measurement signal gets progressively more difficult. On the other hand, increasing $T_m$ in the pulsed regime is possible by increasing the QND interaction rate $g\alpha$ [Fig.~\ref{fig:pulsed}(c)]. This advantage of the pulsed readout is offset by the reduced signal transfer coefficient $T_s$ and reduced measurement-induced squeezing due to dissipation during the readout. More general trade-offs between the transfer coefficients and conditional variance might be obtained with different pulse shape functions $f_{\rm out}$; the $TV$ diagram provides an ideal tool for analyzing such trade-offs and their consequences.

		\section{Discussion and conclusions}\label{sec:summary}
		
		Our work presents a comprehensive formalism for quantifying the performance of optomechanical measurements using relevant figures of merit inspired by quantum optics---the conditional variance of the mechanical quadrature and transfer coefficients characterizing how well the signal is preserved in the mechanical and optical modes. We have shown the application of this approach to study various measurement scenarios, using our framework to derive known results within and obtain new insights into existing measurement strategies and their limitations. Specifically, we confirmed that displacement detection is a fully classical measurement with optimal performance at the standard quantum limit, which minimizes the conditional variance of the mechanical mode. Coherent quantum noise cancellation is, surprisingly, also a classical measurement, as long as detection is performed on mechanical resonance; optimizing detection frequency allows improving the measurement and reaching the quantum nondemolition (QND) regime. Finally, for single-quadrature measurements, we derived a simple and experimentally accessible condition on optomechanical cooperativity to achieve the QND regime.
		
		Furthermore, we studied the impact of various experimental imperfections in the form of fast oscillating optomechanical interaction terms (typically neglected under the rotating wave approximation) and additional static bilinear Hamiltonians. Within our framework, the reduced measurement efficiency can be characterized by a generalized standard quantum limit which can be obtained in a straightforward manner as the minimum achievable conditional variance of the mechanical quadrature of interest. This approach is relevant especially for the treatment of counterrotating terms and free oscillations and squeezing of the mechanical mode, which lead to contamination of the measured mechanical quadrature with backaction noise. Crucially, we showed that it is possible to use these imperfections to compensate each other, for example, by intentionally introducing mechanical detuning to offset the effect of mechanical squeezing and regain perfect QND measurement.
		
		Thanks to these insights, we were then able to propose an approach to QND measurements in levitodynamics using coherent scattering. Modulating the tweezer trapping a nanoparticle at the mechanical frequency leads to a strong single-quadrature interaction reminiscent of QND readout in conventional optomechanics. However, coherent scattering between the mechanical sidebands creates and annihilates pairs of phonons, introducing squeezing which mixes both mechanical quadratures and thus backaction noise to the quadrature of interest. Suitable detuning of the modulation frequency from the mechanical sidebands can then be used to obtain true QND measurement, opening new avenues for tests of fundamental physics and quantum technology applications in levitodynamics. To find the optimal regime for QND measurements, we studied two distinct scenarios: In the first one, the readout is applied simultaneously with another tweezer used for preparing a quantum state of interest, while in the second, state preparation and measurement are executed sequentially. While the latter approach suffers from faster thermalization of the mechanical state, it allows a stronger measurement since only the readout tweezer is being used. Using our formalism, we investigated the trade-offs associated with this choice and identified regimes where each strategy is advantageous.
		
		To fully benefit from the formalism developed in this article, the relevant figures of merit---conditional mechanical variance and signal and meter transfer coefficients---should be directly accessible from measurement data. The approach discussed here allows them to be evaluated when the full system scattering matrix and initial covariances are known, which can all be measured or inferred. However, this requires full characterization of the optomechanical system which is resource intensive. An important open question is whether these quantities can be determined in a simpler procedure requiring fewer measurements than a full system tomography. This would then allow efficient evaluation of optomechanical measurements and open the way to their further optimization.

		\begin{acknowledgments}
			This work has been funded by a grant from the Programme Johannes Amos Comenius under the Ministry of Education, Youth and Sports of the Czech Republic CZ.02.01.01/00/22\_008/0004649  of MEYS \v{C}R. As set out in the Legal Act, beneficiaries must ensure that the open access to the published version or the final peer-reviewed manuscript accepted for publication is provided immediately after the date of publication via a trusted repository under the latest available version of the Creative Commons Attribution International Public Licence (CC BY) or a licence with equivalent rights. For long-text formats, CC BY-NC, CC BY-ND, CC BY-NC-ND or equivalent licenses could be applied.
			In addition, O.\v{C} was supported by the project 8J21AT007 of MEYS \v{C}R;
			F.B.and R. F. acknowledge the support of project 23-06308S of the Czech Science Foundation and by Project No. LUC25006 of MEYS Czech Republic.
		\end{acknowledgments}

		\appendix
		\section{System dynamics with displacement detection}\label{app:efficiency}
		
		To obtain the scattering matrix for displacement detection, we start by writing the equations of motion~\eqref{eq:EOM} in the matrix form
		\begin{equation}\label{eq:EOM_compact}
			\dot{\bf u} = {\bf Au} + {\bf Hu}\inpt,
		\end{equation}
		where ${\bf u} = (X,Y,x,p)^T$ and we introduced the drift matrix
		\begin{equation}
			{\bf A} = \begin{pmatrix}
				-\frac{\kappa}{2} & 0 & 0 & 0 \\
				0 & -\frac{\kappa}{2} & -2g & 0 \\
				0 & 0 & -\frac{\gamma}{2} & \omega_m \\
				-2g & 0 & -\omega_m & -\frac{\gamma}{2}
			\end{pmatrix}
		\end{equation}
		and ${\bf H} = \diag(\sqrt{\kappa},\sqrt{\kappa},\sqrt{\gamma},\sqrt{\gamma})$. Combined with the input--output relation, ${\bf u}\out = \bf{Hu} - {\bf u}\inpt$, the equations of motion can be solved in the frequency space defined via the Fourier transform
		\begin{equation}
			o(\omega) = \int_{-\infty}^\infty o(t)e^{i\omega t} dt.
		\end{equation}
		We thus obtain the relation between input and output fields,
		\begin{equation}
			{\bf u}\out(\omega) = -[{\bf H}({\bf A}+i\omega {\bf I}_4)^{-1} {\bf H} + {\bf I}_4]{\bf u}\inpt(\omega) = {\bf S}(\omega){\bf u}\inpt(\omega),
		\end{equation}
		where ${\bf I}_4$ is the $4\times 4$ identity matrix and
		\begin{equation}\label{eq:app_scatter}
			{\bf S}(\omega) = \begin{pmatrix}
				K & 0 & 0 & 0 \\
				t_m\chi_a^2\tilde{\chi}_b & K & -s\chi_a\mu_b & -u_m\chi_a\tilde{\chi}_b \\
				-u_m\chi_a\tilde{\chi}_b & 0 & \tilde{\Gamma} & \tilde{\Omega} \\
				-s\chi_a\mu_b & 0 & -\tilde{\Omega} & \tilde{\Gamma}
			\end{pmatrix}
		\end{equation}
		is the scattering matrix. Here, we defined 
		\begin{subequations}\label{eq:app_def}
			\begin{align}
				\chi_a(\omega) &= (\kappa-2i\omega)^{-1}, \\
				\tilde\chi_b(\omega) &= [(\gamma-2i\omega)^2+4\omega_m^2]^{-1}, \\
				\mu_b(\omega) &= \frac{\gamma-2i\omega}{(\gamma-2i\omega)^2+4\omega_m^2},\\
				K(\omega) &= \frac{\kappa+2i\omega}{\kappa-2i\omega}, \\
				\tilde{\Gamma}(\omega) &= \frac{\gamma^2+4\omega^2-4\omega_m^2}{(\gamma-2i\omega)^2+4\omega_m^2},\\
				\tilde{\Omega}(\omega) &= \frac{4\gamma\omega_m}{(\gamma-2i\omega)^2+4\omega_m^2},\\
				s &= 4\sqrt{C}\kappa\gamma,\\
				t_m &= 16C\kappa^2\gamma\omega_m,\\
				u_m &= 8\sqrt{C}\kappa\gamma\omega_m;
			\end{align}
		\end{subequations}
		note that we do not include the explicit frequency dependence of these quantities in the scattering matrix for brevity.
		
		The scattering matrix can be used to obtain the conditional variance and transfer coefficients. Assuming the input covariance matrix as given in Eq.~\eqref{eq:Vin}, we get the following elements of the output covariance matrix:
		\begin{subequations}
			\begin{align}
				\begin{split}
					V\out^{22} &= |S_{23}|^2V_x + 2S_{24}S_{23}V_{xp} + |S_{24}|^2V_p \\
					&\quad + \frac{1}{2}(|S_{21}|^2 + |S_{22}|^2), 
				\end{split} \\
				\begin{split}
					V\out^{33} &= |S_{33}|^2V_x + 2S_{33}S_{34}V_{xp} + |S_{34}|^2V_p \\
					&\quad + \frac{1}{2}(|S_{31}|^2 + |S_{32}|^2),
				\end{split} \\
				\begin{split}
					V\out^{32} &= S_{23}S_{33}V_x + (S_{24}S_{33} + S_{23}S_{34})V_{xp} \\
					&\quad + S_{24}S_{34}V_p + \frac{1}{2}(S_{21}S_{31} + S_{22}S_{32}),
				\end{split}
			\end{align}
		\end{subequations}
		where $V\out^{33}$ and $V\out^{22}$ denote the variances of the mechanical position and optical phase quadratures, respectively, and $V\out^{32}$ is their cross-covariance. From these, we evaluate the conditional variance of the mechanical quadrature after an optical measurement, $V_c = V\out^{33}-|V\out^{32}|^2/V\out^{22}$, and the measurement-equivalent noises~\cite{Roch1992}
		\begin{subequations}
			\begin{align} 
				n_s^{\rm eq} &= \frac{2S_{33}S_{34}}{|S_{33}|^2}V_{xp} + \frac{|S_{34}|^2}{|S_{33}|^2}V_p + \frac{1}{2}\frac{|S_{31}|^2+|S_{32}|^2}{|S_{33}|^2}, \label{Eq:s_noise}\\
				n_m^{\rm eq} &= \frac{2S_{24}S_{23}}{|S_{23}|^2}V_{xp} + \frac{|S_{24}|^2}{|S_{23}|^2}V_p + \frac{1}{2}\frac{|S_{21}|^2+|S_{22}|^2}{|S_{23}|^2}.  \label{Eq:m_noise}
			\end{align}	
		\end{subequations}
		Here, we defined the initial mechanical variances $V_x = (n_m + \Re m_{\rm sq} + \frac{1}{2})$, $V_p = (n_m - \Re m_{\rm sq} + \frac{1}{2})$, and their correlation $V_{xp} = \Im m_{\rm sq}$.
		
		From the scattering matrix, one can obtain the conditional variance and transfer coefficients in a straightforward manner; however, since the final expressions are complex and do not offer any insight into the dynamics, we omit them here and show only the graphical solution in Fig.~\ref{fig:displ}. Moreover, the scatterig matrix can be used to determine the cooperativity corresponding to the SQL: Equating the imprecision and backaction noise in the cavity output, $|S_{21}(\omega)| = |S_{22}(\omega)|$, yields the cooperativity
		\begin{equation}
			C_{\rm SQL} = \frac{\kappa^2 + 4\omega^2}{16\kappa^2\gamma\omega_m}|4\omega_m^2 + (2i\omega  - \gamma)^2|.
		\end{equation}
		For detection on mechanical resonance, $\omega = \omega_m$, and a high-quality mechanical mode, $\gamma\ll\omega_m$, this expression simplifies to $C_{\rm SQL}\approx \frac{1}{4}+\omega_m^2/\kappa^2$.
		
		To account for losses in detection, we further modify the optical output quadratures according to
		\begin{equation}
			R\out(t) \to \sqrt{\eta}R\out(t) + \sqrt{1-\eta}R\inpt^\eta(t),
			\label{Eq:detection_loss}
		\end{equation}
		where $R = X,Y$, $0\leq\eta\leq 1$ is the detection efficiency, and the operators $R\inpt^\eta$ account for the additional fluctuations due to loss and have the same correlation functions as $R\inpt$. We can then define a $4\times6$ scattering matrix to include the additional mode with
\begin{widetext}
\begin{equation}\label{eq:app_scatter}
	{\bf S}(\omega) = \begin{pmatrix}
		\sqrt \eta K & 0 & 0 & 0 &\sqrt {1 - \eta }&0\\
		\sqrt \eta t_m\chi_a^2\tilde{\chi}_b & \sqrt \eta K & -\sqrt \eta s\chi_a\mu_b & -\sqrt \eta u_m\chi_a\tilde{\chi}_b &0&\sqrt {1 - \eta }\\
		-u_m\chi_a\tilde{\chi}_b & 0 & \tilde{\Gamma} & \tilde{\Omega} &0&0\\
		-s\chi_a\mu_b & 0 & -\tilde{\Omega} & \tilde{\Gamma}&0&0
	\end{pmatrix}
\end{equation}
\end{widetext}
		The scattering matrix now connects the vector of input operators ${\bf u}\inpt = (X\inpt,Y\inpt,x\inpt,p\inpt,X\inpt^\eta,Y\inpt^\eta)^T$ to the output ${\bf u}\out = (X\out,Y\out,x\out,p\out)^T$. Although we could include the output quadratures for the loss mode, $R_{\rm out}^\eta$, this mode is not accessible for measurement so we omit it from the scattering matrix. Consequently, in the measurement-equivalent noise expressions of Eqs.~(\ref{Eq:s_noise}-\ref{Eq:m_noise}), the signal noise $n_s$ remains unchanged, while the meter noise $n_m$ renormalizes via $C \to C\eta$. Similarly, the covariance matrix elements $V_{\text{out}}^{33}$, $V_{\text{out}}^{34}$, $V_{\text{out}}^{43}$, and $V_{\text{out}}^{44}$ remain unaffected, whereas all other elements rescale according to $C \to C\eta$.

		\section{Characterization of coherent quantum noise cancellation}\label{app:cqnc}
		
		To evaluate the conditional variance and transfer coefficients for CQNC, we start from the equations of motion corresponding to the CQNC Hamiltonian~\eqref{eq:H_CQNC},
		\begin{equation}
			\dot{\bf u} = {\bf Au}+{\bf Hu}\inpt,
		\end{equation}
		where the quadrature vector ${\bf u} = (X,Y,x,p,X_c,Y_c)^T$ has been expanded to include the quadratures of the negative-mass oscillator, $X_c = (c+c^\dagger)/\sqrt{2}$, $Y_c = -i(c-c^\dagger)/\sqrt{2}$ (and analogous expansions hold for ${\bf u}_{\rm in,out}$). The drift matrix is given by\newline
		\begin{equation}
			{\bf A} = \begin{pmatrix}
				-\frac{\kappa}{2} & 0 & 0 & 0 & 0 & 0 \\
				0 & -\frac{\kappa}{2} & -2g & 0 & -2g & 0 \\
				0 & 0 & -\frac{\gamma}{2} & \omega_m & 0 & 0 \\
				-2g & 0 & -\omega_m & -\frac{\gamma}{2} & 0 & 0 \\
				0 & 0 & 0 & 0 & -\frac{\gamma}{2} & -\omega_m \\
				-2g & 0 & 0 & 0 & \omega_m & -\frac{\gamma}{2}
			\end{pmatrix}
		\end{equation}
		and ${\bf H} = \diag(\sqrt{\kappa},\sqrt{\kappa},\sqrt{\gamma},\sqrt{\gamma},\sqrt{\gamma},\sqrt{\gamma})$. A straightforward calculation then yields the scattering matrix
		\begin{widetext}
			\begin{equation}
				{\bf S}(\omega) = \begin{pmatrix}
					K & 0 & 0 & 0 & 0 & 0 \\
					0 & K & -s\chi_a\mu_b & -u_m\chi_a\tilde{\chi}_b & -s\chi_a\mu_b & u_m\chi_a\tilde{\chi}_b \\
					-u_m\chi_a\tilde{\chi}_b & 0 & \tilde{\Gamma} & \tilde{\Omega} & 0 & 0 \\
					-s\chi_a\mu_b & 0 & -\tilde{\Omega} & \tilde{\Gamma} & 0 & 0 \\
					u_m\chi_a\tilde{\chi}_b & 0 & 0 & 0 & \tilde{\Gamma} & \tilde{\Omega} \\
					-s\chi_a\mu_b & 0 & 0 & 0 & -\tilde{\Omega} & \tilde{\Gamma}
				\end{pmatrix}.
			\end{equation}
		\end{widetext}
		We can see that $S_{21}=0$ which reflects efficient suppression of measurement backaction in the optical output; consequently, nonzero value of the meter transfer coefficient $T_m$ is possible in the strong cooperativity regime. On the other hand, since $S_{31}\neq 0$ in general, measurement backaction still affects the mechanical position quadrature and limits the signal transfer coefficient $T_s$ to values close to zero.
		
		Via their interaction with a common optical cavity mode, the mechanical resonator and the negative-mass oscillator become correlated~\cite{Polzik2015,Huang2018}. The resulting two-mode squeezing between these two modes then manifests as strong thermal noise in the mechanical mode in the position variance conditioned on the optical cavity output, Eq.~\eqref{eq:Vc} (i.e., with the negative-mass oscillator traced out). The correlations can be taken into account by conditioning also on the state of the negative-mass oscillator, which allows the conditional mechanical variance to remain small. To see which of the two quadratures of the negative-mass oscillator is more strongly correlated with the mechanical position quadrature, we evaluate the relevant elements of the covariance matrix,
		\begin{subequations}
			\begin{align}
				V\out^{35} &=  - \frac{32C\gamma^2\kappa^2\omega_m^2}{(\kappa^2 + 4\omega^2)[(\gamma^2 + 4\omega^2)^2 + 8\omega_m^2(\gamma^2 - 4\omega^2 + 2\omega_m^2)]}, \\
				V\out^{36} &= \frac{16C\gamma^3\kappa^2\omega_m}{(\kappa^2 + 4\omega^2)[(\gamma^2 + 4\omega^2)^2 + 8\omega_m^2(\gamma^2 - 4\omega^2 + 2\omega_m^2)]}.
			\end{align}
		\end{subequations}
		Taking their ratio, we obtain
		\begin{equation}
			\frac{V\out^{35}}{V\out^{36}} = -2\frac{\omega_m}{\gamma} = -2Q_m.
		\end{equation}
		Since we assume a high-quality mechanical mode, $\omega_m/\gamma = Q_m\gg 1$, correlations with the amplitude quadrature of the negative-mass oscillator are stronger, and we define the conditional mechanical variance as
		\begin{equation}
			{V_c} = V\out^{33} - \frac{V\out^{22}|V\out^{35}|^2 + |V\out^{23}|^2 V\out^{55} - 2V\out^{23}V\out^{25}V\out^{35}}{V\out^{22}V\out^{55}-|V\out^{25}|^2 }\,,
		\end{equation}
		which accounts for conditioning on both the phase quadrature of the cavity output and the amplitude quadrature of the negative-mass oscillator. We have checked that correlation between negative-mass oscillator and the optical mode, $V\out^{25}$, is much smaller than $V\out^{22}$ and $V\out^{55}$, therefore we recover Eq.~(\ref{eq:Vc_CQNC}) of the main text. The ideal results (i.e., minimum conditional variance) would be obtained by finding the ideal rotated quadrature of the negative-mass oscillator; however, given the large ratio $V\out^{35}/V\out^{36}$, such optimization would provide little improvement.

		\section{Detection losses and thermal cavity noise in QND measurements}\label{app:QND}
		
		For QND measurement, the scattering matrix simplifies by setting $\omega_m = 0$ in Eq.~\eqref{eq:app_scatter}.  Incorporating detection losses following the procedure in Eq.~\eqref{Eq:detection_loss}, the scattering matrix becomes:
		\begin{equation}\label{eq:QND_S}
			{\bf S}(\omega) = \begin{pmatrix}
				\sqrt{\eta}K & 0 & 0 & 0 & \sqrt{1-\eta} & 0 \\
				0 & \sqrt{\eta}K & -\sqrt{\eta}s\chi_a\chi_b & 0 & 0 & \sqrt{1-\eta} \\
				0 & 0 & \Gamma & 0 & 0 & 0 \\
				-s\chi_a\chi_b & 0 & 0 & \Gamma & 0 & 0
			\end{pmatrix},
		\end{equation}	
		where $\chi_b(\omega) = (\gamma-2i\omega)^{-1}$ and the remaining quantities are defined in Eqs. \eqref{eq:app_def}.
		
		Finally, to account for finite temperature of the cavity input, we extend the correlation functions, Eq.~\eqref{eq:cav_in}, to
		\begin{subequations}
			\begin{align}
				\langle X_{\rm in}(t)X_{\rm in}(t')\rangle  &= \langle Y_{\rm in}(t)Y_{\rm in}(t')\rangle  = \left(n_c + \frac{1}{2}\right)\delta (t - t'),\\
				\langle X_{\rm in}(t){Y_{\rm in}}(t')\rangle  &=  - \langle Y_{\rm in}(t)X_{\rm in}(t')\rangle  =   \frac{i}{2}\delta (t - t'),
			\end{align}
		\end{subequations}
		where $n_c$ is the average occupation of the cavity bath (with the same modifications for $R\inpt^\eta$). With these modifications, the conditional variance and transfer coefficients become
		\begin{subequations}
			\begin{align}
				V_c &= \frac{1}{16C\eta(n_c+\frac{1}{2})^{-1} + V_x^{-1}}, \\
				T_s &= 1, \\
				T_m &= \frac{16C\eta(n_c+\frac{1}{2})^{-1}}{16C\eta(n_c+\frac{1}{2})^{-1}+V_x^{-1}},
			\end{align}
		\end{subequations}
		which correspond to the expressions in the main text [Eqs. \eqref{eq:QNDIdeal}] with the replacement $C\to C\eta(n_c+\frac{1}{2})^{-1}$.

		\section{QND measurement with systematic imperfections}\label{app:imperfections}

		\subsection{Cavity detuning}
		
		For the QND Hamiltonian with free cavity oscillation, Eq.~\eqref{eq:QND_dc}, we obtain the scattering matrix
		\begin{equation}
			{\bf S}(\omega) = \begin{pmatrix}
				\tilde{K} & \tilde{\Delta} - u_c\tilde{\chi}_a\chi_b & 0 \\
				-\tilde{\Delta} & \tilde{K} & -s\mu_a\chi_b & 0 \\
				0 & 0 & \Gamma & 0 \\
				-s\mu_a\chi_b & -u_c\tilde{\chi}_a\chi_b & t_c\tilde{\chi}_a\chi_b^2 & \Gamma
			\end{pmatrix},
		\end{equation}
		where
		\begin{subequations}
			\begin{align}
				\tilde{\chi}_a(\omega) &= [(\kappa-2i\omega)^2 + 4\delta_c^2]^{-1},\\
				\tilde{K}(\omega) &= \frac{\kappa^2+4\omega^2-4\delta_c^2}{(\kappa-2i\omega)^2 + 4\delta_c^2},\\
				\tilde{\Delta}(\omega) &= \frac{4\kappa\delta_c}{(\kappa-2i\omega)^2 + 4\delta_c^2},\\
				\Gamma(\omega) &= \frac{\gamma+2i\omega}{\gamma-2i\omega},\\
				t_c &= 16C\kappa\gamma^2\delta_c,\\
				u_c &= 8\sqrt{C}\kappa\gamma\delta_c.
			\end{align}
		\end{subequations}
		From the scattering matrix, we can directly obtain the conditional variance and transfer coefficients,
		\begin{subequations}
			\begin{align}
				V_c &= \frac{1}{V_x^{-1} + 32C\kappa^4/(\kappa^2+4\delta_c^2)^2},\\
				T_s &= 1, \\
				T_m &= \frac{32C\kappa^4/(\kappa^2+4\delta_c^2)^2}{V_x^{-1} + 32C\kappa^4/(\kappa^2+4\delta_c^2)^2}.
			\end{align}
		\end{subequations}
		These expressions are identical to the case of ideal QND readout up to the cooperativity rescaled by the factor $\kappa^4/(\kappa^2+4\delta_c^2)^2$; nonzero detuning of the cavity field from the exact resonance therefore limits the detection efficiency.

		\subsection{Mechanical dynamics}
		
		For the Hamiltonian with the quadratic momentum term,
		\begin{equation}
			H = \nu p^2 + 2gXx,
		\end{equation}
		a straightforward calculation gives the scattering matrix
		\begin{equation}
			{\bf S}(\omega) = \begin{pmatrix}
				{K(\omega )}&0&0&0\\
				{{t_\nu }\chi _b^2\chi _a^2}&{K(\omega )}&{ - s{\chi _b}{\chi _a}}&{ - {u_\nu }\chi _b^2{\chi _a}}\\
				{ - {u_\nu }\chi _b^2{\chi _a}}&0&{\Gamma (\omega )}&{8\gamma \nu \chi _b^2}\\
				{ - 4s{\chi _b}{\chi _a}}&0&0&{\Gamma (\omega )}
			\end{pmatrix}
		\end{equation}
		where
		\begin{subequations}
			\begin{align}
				{t_\nu } =& 32C\gamma {\kappa ^2}\nu \\
				{u_\nu } =& 16\sqrt C \gamma \kappa \nu 
			\end{align}
		\end{subequations}
		Moreover, we obtain the conditional variance
		\vspace{0.1in}
		\begin{widetext}
			\begin{equation}
				V_c=\frac{{{\gamma ^2}{V_x} + 16\gamma \nu {V_{xp}} + 64{\nu ^2}\left[ {\left( {2C + {V_p}} \right)\left( {1 + 8C{V_x}} \right) - 8CV_{xp}^2} \right]}}{{512C{\nu ^2}\left( {2C + {V_p}} \right) + {\gamma ^2}\left( {1 + 32C{V_x}} \right) + 256C\gamma \nu {V_{xp}}}}\,,
			\end{equation}
		\end{widetext}
		and transfer coefficients
		\begin{subequations}
			\begin{align}
				T_s &= \frac{{{V_x}{\gamma ^2}}}{{{V_x}{\gamma ^2} + 16\nu \left[ {4\nu \left( {2C + {V_p}} \right) + \gamma {V_{xp}}} \right]}}\,,\\
				T_m &= \frac{{32C{\gamma ^2}{V_x}}}{{512C{\nu ^2}\left( {2C + {V_p}} \right) + {\gamma ^2}\left( {1 + 32C{V_x}} \right) + 256C\gamma \nu {V_{xp}}}}\,,
			\end{align}
		\end{subequations}
		where we assume detection on cavity resonance, $\omega = 0$.
		Clearly, the conditional variance and transfer coefficients depend not only on the initial position variance, $V_x = n_m+\Re{}m+\frac{1}{2}$, but also on the initial momentum variance, $V_p = n_m-\Re m_{\rm sq}+\frac{1}{2}$ and the correlations between the two, $V_{xp}=\Im m_{\rm sq}$. This is a direct consequence of the equation of motion~\eqref{eq:EOM_x}, which mixes both mechanical quadratures, making them both visible in the optical and mechanical outputs $Y\out, x\out$. Remarkably, the transfer coefficients can reach values beyond unity, $T_{s,m} > 1$, when the position and momentum quadratures are anti-correlated, $\Im m_{\rm sq} < 0$. This situation arises always when the measurement device mixes the input quadratures and these exhibit initial correlations and results in negative measurement-equivalent input noises~\cite{Roch1992}.
		
		Finally, for the Hamiltonian
		\begin{equation}
			H = \frac{\xi}{2}(xp+px)+2gXx,
		\end{equation}
		we get the scattering matrix
		\begin{equation}
			{\bf S}(\omega) = \begin{pmatrix}
				K&0&0&0\\
				0&K&{ - s{\chi _a}{\mu _ - }}&0\\
				0&0&{{\Gamma _ + }}&0\\
				{ - s{\chi _a}{\mu _ + }}&0&0&{{\Gamma _ - }}
			\end{pmatrix},
		\end{equation}
		where
		\begin{subequations}
			\begin{align}
				\mu _ \pm =& {\left( {\gamma  \pm 2\zeta  - 2i\omega } \right)^{ - 1}}\,,\\
				\Gamma _ \pm  =& \frac{{\gamma  \pm 2\zeta  + i\omega }}{{\gamma  \mp 2\zeta  - i\omega }}\,.
			\end{align}
		\end{subequations}
		This matrix has the same structure as the scattering matrix for ideal QND readout [cf. Eq.~\eqref{eq:QND_S} in the limit $\eta = 1$] and so the behavior of the system remains qualitatively the same. The different expressions for the various nonzero elements of the scattering matrix only result in quantitative differences in the conditional variance and transfer coefficients as given in Eqs.~\eqref{eq:VTT_xi} and discussed in the main text.

		\section{Floquet analysis of counterrotating terms}\label{app:Floquet}
		
		The fast-oscillating terms ignored under the RWA are crucial for fully understanding and evaluating the quality of an optomechanical QND measurements with finite sideband ratio. To understand the effect of these counter-rotating terms, we consider the Hamiltonian  
		\begin{equation}
			H = 2gXx + \sqrt{2} gX (b e^{2i\omega_m t} + b^\dag e^{- 2i\omega_m t}).
		\end{equation}
		We then apply the quantum Langevin approach to derive the system evolution as
		\begin{equation}\label{compact_matrix_form_EQM2}
			\dot{\bf{u}}(t) = {\bf A}(t){\bf u} (t)+ {\bf{H u}}_{{\rm{in}}}(t)
		\end{equation}
		with the time-dependent drift matrix
		\begin{equation}
			{\bf A}(t) = \begin{pmatrix}
				-\frac{\kappa}{2} & 0 & 0 & 0 \\
				0 & -\frac{\kappa}{2} & -2g(1 + c) & 2gs \\
				-2gs & 0 & -\frac{\gamma}{2} & 0 \\
				-2g(1+c) & 0 & 0 & -\frac{\gamma}{2}
			\end{pmatrix},
		\end{equation}
		where $c = \cos(2\omega_m t)$, $s = \sin(2\omega_m t)$.
		
		We study the resulting system dynamics by employing the Floquet approach~\cite{Malz2016,Malz2016a}. We write the time-dependent drift matrix in its frequency components as ${\bf A}(t) = {\bf A}^{(-1)} e^{-2i\omega_m t} + {\bf A}^{(0)} + {\bf A}^{(1)} e^{2i\omega_m t}$, where the individual matrices ${\bf A}^{(n)}$ are time independent,
		\begin{subequations}
			\begin{align} 
				{\bf A}^{(-1)} &= \begin{pmatrix}
					0 & 0 & 0 & 0 \\
					0 & 0 & -g & ig \\
					-ig & 0 & 0 & 0 \\
					-g & 0 & 0 & 0
				\end{pmatrix}, \\
				{\bf A}^{(0)} &= \begin{pmatrix}
					-\frac{\kappa}{2} & 0 & 0 \\
					0 & -\frac{\kappa}{2} & -2g & 0 \\
					0 & 0 & -\frac{\gamma}{2} & 0 \\
					-2g & 0 & 0 & -\frac{\gamma}{2}
				\end{pmatrix}, \\
				{\bf A}^{(1)} &= \begin{pmatrix}
					0 & 0 & 0 & 0 \\
					0 & 0 & -g & -ig \\
					ig & 0 & 0 & 0 \\
					-g & 0 & 0 & 0
				\end{pmatrix}.
			\end{align}
		\end{subequations}
		We see that the mechanical position quadrature is only affected by the optical amplitude quadrature off resonance, which is highly suppressed by the filtering effect of the cavity. The mechanical momentum quadrature is affected by both on- and off-resonance optical amplitude quadrature, where the cavity resonantly enhances the on-resonance component. 	By defining the quadratures in terms of their Fourier components in a similar manner,
		\begin{equation}
			{\bf u}(t) = \sum_{n = -\infty}^\infty {\bf u}^{(n)}(t) e^{2in\omega_m t},
		\end{equation}
		we can write Eq.~(\ref{compact_matrix_form_EQM2}) in the frequency domain as
		\begin{align}\label{equations_of_motion_BRWA_total}
			\begin{split}
				&-({\bf A}^{(0)}+i\omega{\bf I}_4 - 2in\omega_m{\bf I}_4){\bf u}^{(n)}(\omega) = \\
				&\  {\bf A}^{(-1)}{\bf u}^{(n+1)}(\omega) + {\bf A}^{(1)}{\bf u}^{(n-1)}(\omega) + {\bf H}{\bf u}\inpt(\omega)\delta_{n,0},
			\end{split}
		\end{align}
		where the input noise operators ${\bf u}\inpt(\omega)$ contribute only to the zero-frequency component in the Floquet space, ${\bf u}^{(0)}(\omega)$. We truncate this frequency expantion by considering $ n=\{-1, 0 , 1\} $ to find the approximate solution for the $n = \pm 1$ components,
		\begin{subequations}
			\begin{align}
				{\bf u}^{(-1)}(\omega) &= -({\bf A}^{(0)} + i\omega{\bf I}_4 + 2i\omega_m{\bf I}_4)^{-1} {\bf A}^{(-1)}{\bf u}^{(0)}(\omega), \\
				{\bf u}^{(1)}(\omega) &= -({\bf A}^{(0)} + i\omega{\bf I}_4 - 2i\omega_m{\bf I}_4)^{-1} {\bf A}^{(1)} {\bf u}^{(0)}(\omega),
			\end{align}
		\end{subequations}
		which can be plugged into the equation for the on-resonance component,
		\begin{align}\label{Frequency_domain_2}
			\begin{split}
				& -({\bf A}^{(0)} + i\omega{\bf I}_4) {\bf u}^{(0)}(\omega) = \\ 
				&\qquad ={\bf A}^{(-1)}{\bf u}^{(1)}(\omega) + {\bf A}^{(1)}{\bf u}^{(-1)}(\omega) + {\bf H}{\bf u}\inpt(\omega).
			\end{split}
		\end{align}
		This equation can now be solved to obtain the relationship between the intracavity Floquet component ${\bf u}^{(0)}$ and the input noise ${\bf u}\inpt$,
		\begin{widetext}
			\begin{equation}
				{\bf u}^{(0)}(\omega) = -[{\bf A}^{(0)}+i\omega{\bf I}_4 + {\bf A}^{(-1)}({\bf A}^{(0)}+i\omega{\bf I}_4-2i\omega_m{\bf I}_4)^{-1}{\bf A}^{(1)} - {\bf A}^{(1)}({\bf A}^{(0)}+i\omega{\bf I}_4 +2i\omega_m{\bf I}_4)^{-1}{\bf A}^{(-1)}]^{-1}{\bf Hu}\inpt(\omega).
			\end{equation}	
		\end{widetext}
		To obtain the scattering matrix that connects the input ${\bf u}\inpt$ to the output ${\bf u}\out$, we first write the output fields in terms of their Floquet components
		\begin{equation}\label{eq:FloquetOut}
			{\bf u}\out(t) = \sum_n {\bf u}\out^{(n)}(t) e^{2in\omega_m t},
		\end{equation}
		for which we have the following input--output relation (after transforming from temporal to frequency domain):
		\begin{equation}
			{\bf u}\out^{(n)}(\omega) = {\bf Hu}^{(n)}(\omega)-\delta_{n,0}{\bf u}\inpt(\omega).
		\end{equation}
		The output field at frequency $\omega$ is then given by the sum of the appropriate sidebands of all Floquet components,
		\begin{equation}
			{\bf u}\out(\omega) = \sum_n {\bf u}\out^{(n)}(\omega+2n\omega_m),
		\end{equation}
		which follows from the Fourier transform of the output field in the time domain expressed in the Floquet basis, Eq.~\eqref{eq:FloquetOut}.
		A straightforward calculation including $n\in\{-1,0,1\}$ gives the conditional variance and transfer coefficients
		\begin{subequations}
			\begin{align}
				{V_c} &= \frac{{1 + (8C + V_x^{ - 1})\left( {\frac{{4C{\kappa ^2}}}{{{\kappa ^2} + 16\omega _m^2}}} \right)}}{{V_x^{ - 1} + 32C + V_x^{ - 1}\left( {\frac{{32{C^2}{\kappa ^2}}}{{{\kappa ^2} + 16\omega _m^2}}} \right)}},\\
				{T_s} &= \frac{1}{{1 + 8V_x^{ - 1}{{\left( {\frac{{4\kappa {\omega _m}}}{{{\kappa ^2} + 16\omega _m^2}}} \right)}^2}}},\\
				{T_m} &= \frac{{32C}}{{32C + V_x^{ - 1}\left[ {1 + 64{{\left( {\frac{{4C\kappa {\omega _m}}}{{{\kappa ^2} + 16\omega _m^2}}} \right)}^2}} \right]}},
			\end{align}
		\end{subequations}
		where we assumed $\gamma\ll\kappa,\omega_m$. In the limit $\kappa\to 0$ corresponding to perfect RWA, we recover Eqs. \eqref{eq:QNDIdeal} describing ideal QND readout.

		\section{QND measurements via coherent scattering in levitodynamics}\label{app:levitation}
		
		To derive the QND interaction for levitodynamics, we start from the Hamiltonian describing the cavity field, nanoparticle motion, and their interaction (note that the tweezer is a classical field)~\cite{Gonzalez_Ballestero2019},
		\begin{equation}\label{eq:cavity_field_NP_Hamiltonian}
			H = \omega_{\rm cav}a^\dagger a + \frac{p_m^2}{2m} - \frac{1}{2}\alpha_p[{\bf E}_{\rm tw}(x_m,t) + {\bf E}_{\rm cav}(x_m)]^2.
		\end{equation}
		The square of the tweezer field, for a Gaussian trapping beam and small displacements gives
		\begin{equation}
			{\bf E}_{\rm tw}^2(x_m,t) \approx E_0^2(t)\cos^2(\omega_{\rm tw}t)\left(1 - \frac{2x_m^2}{A_x^2 w^2}\right),
		\end{equation}
		where $\omega_{\rm tw}$ is the tweezer frequency, $A_x$ its area, and $w$ its waist; we assume a general time-dependent tweezer amplitude $E_0(t)$ which allows us to treat its modulation, $E_0(t) = E_0[1+\alpha\cos(\Omega t+\phi)]$. We thus obtain the total trapping potential
		\begin{equation}\label{eq:Vtrap}
			V = \frac{1}{2}m\omega_{\rm tr}^2[1+\alpha\cos(\Omega t+\phi)]^2x_m^2
		\end{equation}
		with the trapping frequency
		\begin{equation}\label{eq:trap_frequency}
			\omega_{\rm tr} = \frac{1}{w^2}\sqrt{\frac{2\varepsilon P_0}{\rho \pi c}},
		\end{equation}
		where $\varepsilon$ and $\rho$ are the permittivity and mass density of the nanoparticle and $P_0$ is the tweezer power. We can now expand the square bracket in the potential,
		\begin{equation}
			\begin{split}
				&[1+\alpha\cos(\Omega t+\phi)]^2 = \\
				&\quad =1 + \frac{\alpha^2}{2} + 2\alpha\cos(\Omega t+\phi) + \frac{\alpha^2}{2}\cos(2\Omega t+2\phi),
			\end{split}
		\end{equation}
		which elucidates the renormalization of the mechanical frequency, $\omega_m = \omega_{\rm tw}\sqrt{1+\alpha^2/2}$.
		
		The square of the cavity field $|{\bf E}_{\rm cav}(x_m)|^2$, gives the radiation pressure optomechanical coupling between the nanoparticle and the cavity field and disappears for a nanoparticle positioned at a node of the cavity mode~\cite{Gonzalez_Ballestero2019}. This is the position we assume for the nanoparticle in our case---as that is where the coherent scattering interaction is maximized---and so we neglect this term.
		
		Finally, the cross term ${\bf E}_{\rm tw}(x_m, t)\cdot{\bf E}_{\rm cav}(x_m)$ describes the coherent scattering of the tweezer photons into the cavity~\cite{Gonzalez_Ballestero2019},
		\begin{subequations}\label{eq:HintCS}
			\begin{align}
				H_{\rm int} &=  -G[1+\alpha\cos(\Omega t+\phi)]\cos(\omega_{\rm tw}t) (a + a^\dagger)x_m, \\
				G &= \varepsilon V E_0 \sqrt{\frac{\omega_{\rm cav}}{2\varepsilon _0 V_{\rm cav}}}, \label{eq:CSG}
			\end{align}
		\end{subequations}
		$V$ and $V_{\rm cav}$ are the volumes of the nanoparticle and the cavity mode. We also assume that the polarizations of the tweezer and cavity mode are parallel, which removes coupling of the remaining two translational modes of the nanoparticle (i.e., modes in the plane perpendicular to the cavity axis). After moving to a frame rotating with $\omega_{\rm tw}a^\dagger a$, the interaction Hamiltonian~\eqref{eq:HintCS}, together with the trapping potential \eqref{eq:Vtrap}, kinetic energy of the nanoparticle, and the free cavity oscillations, $(\omega_{\rm cav}-\omega_{\rm tw})a^\dagger a = \Delta a^\dagger a$, gives the total Hamiltonian \eqref{eq:Hlev} with the modulation terms \eqref{eq:modulation} with $g = Gx_{\rm zpf} = G/\sqrt{2m\omega_m}$.
		
		Next, after introducing the mechanical annihilation operator,
		\begin{equation}
			b = \sqrt{\frac{m\omega_m}{2}}\left(x_m + \frac{i}{m\omega_m}p_m\right),
		\end{equation}
		we set $\Delta = 0$ and move to the rotating frame with respect to $\Omega b^\dagger b$. This transformation gives the Hamiltonian
		\begin{widetext}
			\begin{equation}
				\begin{split}
					H &= (\omega_m - \Omega)b^\dagger b + \frac{\omega_m}{4(2+\alpha^2)}\left[2\alpha\cos(\Omega t+\phi)+\frac{\alpha^2}{2}\cos(2\Omega t+2\phi)\right](b^2e^{-2i\Omega t}+b^{\dagger 2}e^{2i\Omega t}) \\
					&\quad - \frac{g}{2}[1+\alpha\cos(\Omega t+\phi)](a+a^\dagger)(be^{-i\Omega t}+b^\dagger e^{i\Omega t}).
				\end{split}
			\end{equation}
		\end{widetext}
		We now apply the rotating wave approximation to get rid of all terms oscillating with $\Omega$ and its multiples,
		\begin{equation}
			\begin{split}
				H &\approx (\omega_m-\Omega)b^\dagger b + \frac{\alpha^2\omega_m}{16(2+\alpha^2)}(b^2e^{2i\phi}+b^\dagger e^{-2i\phi}) \\
				&\quad -\frac{\alpha g}{4}(a+a^\dagger)(be^{i\phi}+b^\dagger e^{-i\phi}).
			\end{split}
		\end{equation}
		Expressing this Hamiltonian in terms of quadrature operators, we get Eq.~\eqref{eq:HQNDlev} of the main text. Nonzero modulation phase, $\phi\neq 0$, then results in rotated quadrature operators for the mechanical mode, $x_\phi = x\cos\phi - p\sin\phi$, $p_\phi = x\sin\phi + p\cos\phi$,
		\begin{equation}
			\begin{split}
				H &= \frac{\omega_m-\Omega}{2}(x_\phi^2+p_\phi^2) +\frac{\alpha^2\omega_m}{16(2+\alpha^2)}(x_\phi^2-p_\phi^2) \\
				&\quad -\frac{1}{2}{\alpha g}Xx_\phi;
			\end{split}
		\end{equation}
		the modulation phase therefore allows the measurement of an arbitrary mechanical quadrature.

		\section{Dual-tweezer levitodynamics}\label{app:dual_tweezer}
		
		Simultaneous state stabilization and readout in levitodynamics requires extending the interaction between the nanoparticle and optical fields, $H_{\rm int} = -\frac{1}{2}\alpha_p|\mathbf{E}(x_m,t)|^2$, to two tweezer beams and two cavity modes,
		\begin{equation}
			\mathbf{E}(x_m,t) = \sum_{i=1}^2 [\mathbf{E}_{\rm tw}^{(i)}(x_m,t)+\mathbf{E}_{\rm cav}^{(i)}(x_m)].
		\end{equation}
		We consider a modulated amplitude for each tweezer $E_0^{(i)}(t) = E_0^{(i)}[1+\alpha_i \cos(\Omega_i t+\phi_i)]$ with generally different modulation depths, frequencies, and phases $\alpha_i,\Omega_i,\phi_i$. We assume that each tweezer is scattered into a different cavity mode; since their frequencies are at least one free spectral range apart (barring small differences due to mechanical sidebands), there will be no scattering between fields with different indices. Neglecting the radiation-pressure interaction as before, we need to evaluate the interaction Hamiltonian
		\begin{equation}
			H_{\rm int} = -\frac{1}{2}\alpha_p\sum_i\left[\mathbf{E}_{\rm tw}^{(i) 2}(x_m,t) + 2\mathbf{E}_{\rm tw}^{(i)}(x_m,t)\cdot\mathbf{E}_{\rm cav}^{(i)}(x_m) \right].
		\end{equation}
		
		Repeating the approach outlined in Appendix~\ref{app:levitation}, we obtain
		\begin{align}
			\begin{split}
				H &= \sum\limits_{i = 1,2} {{\Delta _i}a_i^\dag {a_i}}  + {\omega _m}{b^\dag }b\\
				&+ \sum\limits_{i = 1,2} {\frac{{{\alpha _i}{\omega _{i,\rm tr}^2}}}{{2{\omega _m}}}\left[ {\cos ({\Omega _i}t + {\phi _i}) + \frac{{{\alpha _i}}}{4}\cos 2({\Omega _i}t + \phi )} \right]} (b + b^\dag )^2\\
				&- \sum\limits_{i = 1,2} {\frac{g_i}{2}\left[ {1 +\alpha_i \cos ({\Omega _i}t + {\phi _i})} \right]({a_i} + a_i^\dag )(b + {b^\dag })},
			\end{split}
		\end{align}
		where $\Delta_i = \omega_{i,{\rm cav}} - \omega_{i,{\rm tw}}$, $\omega_m^2 = \sum_i\omega_{i,{\rm tr}}^2(1+\frac{1}{2}\alpha_i^2)$, and $g_i = G_i x_{\rm zpf}$, with $\omega_{{\rm tr},i}$ and $G_i$ given in Eq.~\eqref{eq:trap_frequency} and \eqref{eq:CSG}, respectively. Next, we move to a frame rotating with respect to the Hamiltonian $H_0 = \Omega_2 b^\dagger b+\Delta_1 a_1^\dagger a_1+\Delta_2 a_2^\dagger a_2$ and set $ \Delta_1=\Omega_2 $, $ \Delta_2=0 $, $ \Omega_1 = 2\Omega_2 $. Invoking the rotating wave approximation, we get the system Hamiltonian
		\begin{align}
			\begin{split}
				H &= ({\omega _m} - {\Omega _2}){b^\dag }b + \frac{{{\alpha _1}{\omega _{1,\rm tr}^2}}}{{4{\omega _m}}}\left( {{b^2}{e^{2i{\phi _1}}} + {b^{\dag 2}}{e^{ - 2i{\phi _1}}}} \right)\\
				&+ \frac{{\alpha _2^2{\omega _{2,\rm tr}^2}}}{{16{\omega _m}}}({b^2}{e^{2i{\phi _2}}} + {b^{\dag 2}}{e^{ - 2i{\phi _2}}}) \\
				&- \frac{{{g_1}}}{2}({a_1}{b^\dag } + a_1^\dag b)
				-{\frac{{{g_1}{\alpha _1}}}{4}({a_1}b{e^{i{\phi _1}}} + a_1^\dag {b^\dag }{e^{ - i{\phi _1}}})}\\&- \frac{{{g_2}{\alpha _2}}}{4}({a_2} + a_2^\dag )(b{e^{i{\phi _2}}} + {b^\dag }{e^{ - i{\phi _2}}}).
			\end{split}
		\end{align}
		This Hamiltonian captures the dissipative squeezing dynamics induced by the first cavity mode with nondemolition readout of the mechanical state via the second cavity mode. In the limit where the modulation amplitude of the first tweezer approaches zero $\alpha_1\to 0$, we get sideband cooling of the mechanical motion by its beam-splitter interaction with the first cavity mode.
		
		We now return to the quadrature operators, setting $\phi_1=0, \phi_2=\phi$ without loss of generality,
		to express the system Hamiltonian in the form
		\begin{align}
			\begin{split}
				H &= \frac{{{\omega _m} - {\Omega _2}}}{2}({x^2} + {p^2}) + {{\tilde \alpha }_1}({x^2} - {p^2})\\
				&\quad+ {{\tilde \alpha }_2}\left[ {({x^2} - {p^2})\cos 2\phi  - (xp + px)\sin 2\phi } \right]\\
				&\quad- \frac{{{g_1}}}{2}({X_1}x + {Y_1}p) - \frac{{{g_1}{\alpha _1}}}{4}({X_1}x - {Y_1}p)\\
				&\quad- \frac{{{g_2}{\alpha _2}}}{2}{X_2}(x\cos \phi  - p\sin \phi )
			\end{split}
		\end{align}
		with
		\begin{equation}
			{{\tilde \alpha }_1} = \frac{{{\alpha _1}\omega _{1,{\rm{tr}}}^2}}{{4{\omega _m}}}\,, \qquad
			{{\tilde \alpha }_2} = \frac{{\alpha _2^2\omega _{2,{\rm{tr}}}^2}}{{16{\omega _m}}}\,.
		\end{equation}
		This Hamiltonian combines the parametric and dissipative squeezing generated by the primary tweezer and QND readout due to the second tweezer; this readout is accompanied by detuned parametric squeezing. Defining the quadrature vector ${\bf u} = (X_1,Y_1,X_2,Y_2,x,p)^T$ and the corresponding vector of input and output noises ${\bf u}_{\rm in,out}$ and setting $\phi = 0$ like in the main text for simplicity, we can write the Heisenberg–Langevin equations equations of motion in the matrix form as
		\begin{equation}
			\dot{\bf u} = {\bf Au} + {\bf Hu}\inpt,
		\end{equation}
		where we have defined the drift matrix
		\begin{widetext}
			\begin{equation}
				\mathbf{A} = \begin{pmatrix}
					{ - \frac{{{\kappa _1}}}{2}}&0&0&0&0&{\frac{1}{4}({\alpha _1} - 2){g_1}} \\ 
					0&{ - \frac{{{\kappa _1}}}{2}}&0&0&{\frac{1}{4}({\alpha _1} + 2){g_1}}&0 \\ 
					0&0&{ - \frac{{{\kappa _2}}}{2}}&0&0&0 \\ 
					0&0&0&{ - \frac{{{\kappa _2}}}{2}}&{\frac{{{g_2}{\alpha _2}}}{2}}&0 \\ 
					0&{\frac{1}{4}({\alpha _1} - 2){g_1}}&0&0&{ - \frac{\gamma }{2}}&0 \\ 
					{\frac{1}{4}({\alpha _1} + 2){g_1}}&0&{\frac{{{\alpha _2}{g_2}}}{2}}&0&{ - 4({{\tilde \alpha }_2} + {{\tilde \alpha }_1})}&{ - \frac{\gamma }{2}}
				\end{pmatrix}
			\end{equation}
		\end{widetext}
		and ${\bf H} = \diag(\sqrt{\kappa_1},\sqrt{\kappa_1},\sqrt{\kappa_2},\sqrt{\kappa_2},\sqrt{\gamma},\sqrt{\gamma})$. The equations of motion are transformed into the frequency space to have
		\begin{equation}
			{\bf u}(\omega) =-( {\bf A} + i\omega {\bf I}_6 )^{-1} {\bf H}{\bf u}\inpt (\omega).
		\end{equation} 
		
		The output of the secondary mode, $Y_{2,{\rm out}}$, carries information about the combined mechanical and primary optical baths which can be grouped in the compound signal
		\begin{align}
			\begin{gathered}
				{{\bar x}_{{\text{in}}}} = \frac{1}{{2\sqrt {{\gamma _m}} }}\left[ {{g_1}\left( {{\alpha _1} - 2} \right){\chi _1}\sqrt {{\kappa _1}} {Y_{1,{\text{in}}}} + 2\sqrt \gamma  {x_{{\text{in}}}}} \right]\,,\\
				{{\bar p}_{{\text{in}}}} = \frac{1}{{2\sqrt {{\gamma _m}} }}\left[ {{g_1}\left( {{\alpha _1} + 2} \right){\chi _1}\sqrt {{\kappa _1}} {X_{1,{\text{in}}}} + 2\sqrt \gamma  {p_{{\text{in}}}}} \right]\,,\\ 
			\end{gathered} 
		\end{align}
		where we have defined
		\begin{align}
			\chi_i(\omega)& = \left(\kappa_i - 2i\omega  \right)^{ - 1}\\
			\gamma_m& = \gamma + g_1^2\left( 1 - \frac{\alpha _1^2}{4} \right)/\kappa_1. 
		\end{align}	
		We can now define a reduced vector of quadratures $\bar{\bf u} = (X_2,Y_2,x,p)^T$ and its corresponding noise vector $\bar{\bf u}\inpt = (X_{2,\rm in},Y_{2,\rm in},\bar{x}\inpt,\bar{p}\inpt)^T$. This allows us to express the reduced scattering matrix
		\begin{equation}\label{eq:app_scatter_DT}
			{\mathbf{\bar S}} = \left( {\begin{array}{*{20}{c}}
					{{K_m}}&0&0&0 \\ 
					0&{{K_m}}&{{s_m}}&0 \\ 
					0&0&{{\Gamma _m}}&0 \\ 
					{{s_m}}&0&{{{\tilde \Omega }_m}}&{{\Gamma _m}} 
			\end{array}} \right)\,,
		\end{equation}
		where we defined 
		\begin{subequations}\label{eq:app_def__DT}
			\begin{align}
				{K_m}(\omega ) &= \frac{{{\kappa _2} + 2i\omega }}{{{\kappa _2} - 2i\omega }}{\mkern 1mu} , \hfill \\
				{\Gamma _m}(\omega ) &= \frac{{{\gamma _m} + 2i\omega }}{{{\gamma _m} - 2i\omega }}{\mkern 1mu} , \hfill \\
				{s_m} &= 2{g_2}{\alpha _2}{\chi _2}{\chi _m}\sqrt {{\gamma _m}{\kappa _2}} {\mkern 1mu} , \hfill \\ 
				{{\tilde \Omega }_m} &=  - 16\left( {{{\tilde \alpha }_1} + {{\tilde \alpha }_2}} \right){\gamma _m}\chi _m^2\,.
			\end{align}
		\end{subequations}
		The reduced scattering matrix has a structural form analogous to the scattering matrix characterizing an ideal QND measurement given by Eq.~\eqref{eq:QND_S}. Consequently, despite the differences in the underlying physical mechanisms, the qualitative nature of the system's dynamics remains the same. Therefore, by probing the optical mode after its interaction with the mechanical mode, we gain information about the combined state of the mechanical mode and the first optical mode, as characterized by the conditional variance and transfer matrix for ideal QND measurement.

		\section{Pulsed interaction} \label{app:pulsed_tweezer}
		
		We first use optomechanical interaction with the Hamiltonian 
		\begin{align}\label{eq:pulsed_tweezer}
			H = \frac{{\alpha {\omega _m}}}{{4 + 2{\alpha ^2}}}{(b + {b^\dag })^2}\, - \frac{g}{2}(a{b^\dag } + {a^\dag }b) - \frac{{g\alpha }}{4}(ab + {a^\dag }{b^\dag })\,,
		\end{align}
		to prepare the mechanical state of interest, which can be obtained by solving the corresponding steady-state Lyapunov equation. The mechanical state exhibits correlations similar to those described in Eqs.~(\ref{eq:minput}), with the modification that the thermal occupation of the mode $n_0$ is different from the thermal occupation of the bath $n_m$. Following this state preparation step, the tweezer properties are modified to allow tracking the quantum state of the mechanical mode via the QND Hamiltonian 
		\begin{equation}
			H = \frac{\alpha_2^2\omega_m}{4(2+\alpha_2^2)}x^2 - \frac{1}{2}\alpha_2 g_2 Xx.
		\end{equation}
		We can now express the equations of motion in the matrix form with the drift matrix
		\begin{equation}
			{\bf A}=\begin{pmatrix}
				{ - \frac{\kappa }{2}}&0&0&0\\
				0&{ - \frac{\kappa }{2}}&{\frac{1}{2}{\alpha_2}{g_2}}&0\\
				0&0&{ - \frac{\gamma }{2}}&0\\
				{\frac{1}{2}{\alpha_2}{g_2}}&0&{\nu}&{ - \frac{\gamma }{2}}
			\end{pmatrix},
		\end{equation}
		where $\nu = \alpha_2^2\omega_m/8(2+\alpha_2^2)$.
		
		The solution of the equations of motion is given by formal integration,
		\begin{equation}\label{key}
			{\bf u}(t) = {\bf{M}}(t){\bf{u}}(0) + \int\limits_0^t {ds{\bf{M}}(t - s){\bf{H}}{{\bf u}_{{\rm{in}}}}(s)}\,,
		\end{equation}
		where ${\bf{M}}(t) = \exp ({\bf{A}}t)$ with nonzero elements
		\begin{subequations}
			\begin{align}
				{{\bf{M}}_{11}}(t) &= {{\bf{M}}_{22}}(t) = {e^{ - \frac{1}{2}\kappa t}}\,,\\
				{{\bf{M}}_{33}}(t) &= {{\bf{M}}_{44}}(t) = {e^{ - \frac{1}{2}\gamma t}}\,,\\
				{{\bf{M}}_{23}}(t) &= {{\bf{M}}_{41}}(t) =  \frac{{{\alpha _2}{g_2}}}{{   \kappa -\gamma}}\left( {{e^{ - \frac{1}{2}\gamma t}} - {e^{ - \frac{1}{2}\kappa t}}} \right)\,,\\
				{{\bf{M}}_{43}}(t) &= \nu t{e^{ - \frac{1}{2}\gamma t}}.
			\end{align}
		\end{subequations}
		The proper input and output temporal modes are obtained by filtering the corresponding quadrature operators with suitable mode shape functions~\cite{Rakhubovsky2019}
		\begin{align}
			{\cal O}_{{\rm{in}},{\rm out}} (\tau ) = \int\limits_0^\tau  {ds{f_{{\rm{in}},{\rm out}} }(s){o_{{\rm{in}},{\rm out}}}(s)} \,,
			\label{Eq:in_out_field}
		\end{align}
		where
		\begin{subequations}
			\begin{align}
				{f_{\rm  in} }(s) &= \sqrt {\frac{{\kappa }}{{{{{\cal G}} } - 1}}} {{\bf{M}}_{41}}(\tau-s),\\
				{f_{\rm  out} }(s) &= \sqrt {\frac{{\kappa }}{{{{{\cal G}} } - 1}}}{{\bf{M}}_{23}}(s)
			\end{align}
		\end{subequations}
		and the optomechanical gain
		\begin{align}
			{\cal G} &= 1 + \kappa \int\limits_0^\tau  {ds{\bf{M}}_{41}^2(\tau  - s)} \,.
		\end{align}
		
		Since the phase quadrature $Y$ and position quadrature $x$ are uncoupled from the amplitude and momentum quadratures $X,p$, it is sufficient to consider the former two in the following. Their state at time $\tau$ is given by
		\begin{subequations}
			\begin{align}
				{{\cal Y}_{{\rm{out}}}}(\tau ) &= \sqrt {{\cal G} - 1} x(0) + {{\cal Y}_{{B}}}(\tau )\,,\\
				x(\tau ) &= {{\bf{M}}_{33}}(\tau )x(0) + x_B(\tau)
			\end{align}
		\end{subequations}
		with the noise terms corresponding to the input optical and mechanical quantum fluctuations
		\begin{subequations}
			\begin{align}
				\begin{split}
					{\cal Y}_B(\tau ) &= \sqrt{\kappa}\int_0^\tau dt f_{\rm out}(t){\bf M}_{22}(t)Y(0) - \int\limits_0^\tau  {dt{f_{{\rm{out}}}}(t){Y_{{\rm{in}}}}(t)} \\
					&\quad+ \kappa \int\limits_0^\tau  {ds{Y_{{\rm{in}}}}(s)\int\limits_s^\tau  {dt{f_{{\rm{out}}}}(t){{\bf{M}}_{22}}(t - s)} } \\
					&\quad+ \sqrt {\kappa \gamma } \int\limits_0^\tau  {ds{x_{{\rm{in}}}}(s)\int\limits_s^\tau  {dt{f_{{\rm{out}}}}(t){{\bf{M}}_{23}}(t - s)} }\,,
				\end{split} \\
				x_B(\tau)&=\sqrt{\gamma} \int \limits_0^\tau ds   {{\bf{M}}_{33}}(\tau - s){x_{\rm in}}(s)\,.
			\end{align}
		\end{subequations}
		The relevant elements of the covariance matrix can be evaluated from these expressions to be
		\begin{widetext}
			\begin{subequations}\label{eq:app_covariances_pulsed}
				\begin{align}
					{V_{33}} &= {\bf M}_{33}^2(\tau)V_0 +\gamma V_x\int_0^\tau ds {\bf M}_{33}^2(\tau-s) \\
					{V_{32}} &= V_0 \sqrt {{\cal G} - 1} {{\bf{M}}_{33}}(\tau ) + \gamma \sqrt{\kappa}  {V_x}\int_0^\tau  {ds\int_s^\tau  {dt{f_{{\rm{out}}}}(t){{\bf{M}}_{33}}(\tau  - s){{\bf{M}}_{23}}(t - s)} } \,, \\
					\begin{split}
						{V_{22}} &= \frac{1}{2} + ({\cal G}-1)V_0 +\frac{\kappa}{2}\left(\int_0^\tau ds f_{\rm out}(s){\bf M}_{22}(s)\right)^2
						-\kappa \int_0^\tau ds\int_s^\tau dt f_{\rm out}(s) f_{\rm out}(t) {\bf M}_{22}(t-s) \\
						&\quad+ \frac{\kappa^2}{2}\int_0^\tau  {ds\int_s^\tau dt  {\int_s^\tau  {dt'{f_{{\rm{out}}}}(t){f_{{\rm{out}}}}(t'){{\bf{M}}_{22}}(t' - s){{\bf{M}}_{22}}(t - s)} } } \\
						&\quad+ \kappa \gamma {V_x}\int_0^\tau {ds\int_s^\tau dt  {\int_s^\tau  {dt'{f_{{\rm{out}}}}(t){f_{{\rm{out}}}}(t'){{\bf{M}}_{23}}(t' - s){{\bf{M}}_{23}}(t - s)} } },
					\end{split}
				\end{align}
			\end{subequations}
		\end{widetext}
		where $V_0$ is the input variance of the position quadrature $x_{\rm in}$. From Eqs.~\eqref{eq:app_covariances_pulsed}, we can obtain the conditional variance and transfer coefficients in analogy with all previous calculations.

		%
		
	\end{document}